\documentclass[9pt]{article}

\usepackage[latin1]{inputenc}
\usepackage[english]{babel}

\usepackage{amsmath,amssymb,stmaryrd,amsthm}
\usepackage{wasysym}
\usepackage{subeqnarray}
\usepackage{verbatim}
\usepackage[colorlinks=false,linktocpage]{hyperref}%
\usepackage{multicol}
\usepackage[bottom]{footmisc}
\usepackage{cancel}
\usepackage{units}
\usepackage{pgf}
\usepackage{afterpage}
\usepackage{gensymb}
\usepackage{cancel}
\usepackage{color,bm}
\usepackage{longtable}
\usepackage[titletoc]{appendix}
\usepackage[nottoc]{tocbibind}
\usepackage{cases}
\usepackage{caption}
\usepackage{enumerate}
\usepackage{changes}
\usepackage{graphicx}
\usepackage{subfigure}
\usepackage[square,sort,comma,numbers,super]{natbib}
\usepackage{ulem}

\newcommand{\Eq}{Eq.~}
\newcommand{\Eqs}{Eqs.~}
\newcommand{\Fig}{Fig.~}
\newcommand{\Figs}{Figs.~}
\newcommand{\Tab}{Tab.~}

\begin{document}

\begin{center}
\LARGE{Fluid-solid transition in unsteady, homogeneous, granular shear flows}\\[0.4cm]

\large{Dalila Vescovi, Diego Berzi and Claudio di Prisco}\\[0.4cm]
\normalsize{\textit{~Department of Civil and Environmental Engineering, Politecnico di Milano, 20133 Milano, Italy. \\
E-mail: dalila.vescovi@polimi.it; diego.berzi@polimi.it; claudio.diprisco@polimi.it}}
\end{center}

\vskip 1 cm

\begin{abstract}
Discrete element numerical simulations of unsteady, homogeneous shear flows have been performed by instantly applying a constant shear rate to a random, static, isotropic assembly of identical, soft, frictional spheres at either zero or finite pressure by keeping constant the solid volume fraction until the steady state is reached. If the system is slowly sheared, or, equivalently, if the particles are sufficiently rigid, the granular material exhibits either large or small fluctuations in the evolving pressure, depending whether the average number of contacts per particle (coordination number) is less or larger than a critical value. The amplitude of the pressure fluctuations is rate-dependent when the coordination number is less than the critical and rate-independent otherwise, signatures of fluid-like and solid-like behaviour, respectively. The same critical coordination number has been previously found to represent the minimum value at which rate-independent components of the stresses develop in steady, simple shearing and the jamming transition in isotropic random packings. The observed complex behaviour of the measured pressure in the fluid-solid transition clearly suggests the need for incorporating in a nontrivial way the coordination number, the solid volume fraction, the particle stiffness and the intensity of the particle agitation in constitutive models for the onset and the arrest of granular flows.  
\end{abstract}

%

\section{Introduction}\label{Intro}

Granular materials exhibit a complex mechanical behaviour, even in the case of simple flow conditions, whose study involves interdisciplinary concepts like rheology, plasticity and viscosity.

Depending on both the micro-mechanical properties of the grains (surface friction, collisional inelasticity and contact stiffness) and the macroscopic characteristics of the flow (e.g., velocity and bulk density), different flow regimes exist. Consider, for example, a landslide: immediately after the triggering, it behaves like a solid and a sliding motion takes place; but, if its velocity is sufficiently large, the landsliding evolves into a fluid-like process.

When the system is extremely dense, enduring contacts among grains involved in force chains govern its response, which is mainly rate-independent \cite{chi2012}. In this case, deformations are extremely slow because of the continuous rearrangement of the contact network \cite{how1999} and the granular material behaves like a solid, able to resist finite applied shear stresses without deforming. 

On the other hand, when the particles are widely spaced, force chains are inhibited and collisions dissipate the energy of the system. As a consequence, the medium is strongly agitated, the particles are free to move in all directions and the deformations are rapid \cite{gol2003}. The material response is that of a fluid, that is it yields under shear stress, and stresses are rate-dependent. 

The mechanical response of the system during the solid-fluid transition is still an open question \cite{lud2016}, although several constitutive models have been proposed in the literature to deal at least with steady flow configurations \cite{joh1987,sav1998,chi2012,ves2013,ber2015b,ves2016}.

Molecular dynamics simulations based on Discrete Element Method (DEM) have been successfully employed to study granular systems under  different flow regimes, configurations and geometries (collections of results can be found in \cite{gdr2004} and in \cite{del2007}). Whereas several numerical results in the literature concern steady, shearing granular flows \cite{bab1990,cam2002,dac2005,ji2006,mit2007,est2008,chi2012,chi2013,ves2016}, unsteady conditions have been less investigated.

Sun and Sundaresan \cite{sun2011} carried out DEM simulations of unsteady, homogeneous shear flows of frictional spheres at constant solid volume fraction. They focused on the solid regime, where the stresses are rate-independent, and showed that, in that case, the pressure scales with the product of the particle stiffness and the square of the distance of the coordination number, i.e., the average number of contacts per particle, from a critical value.

Here, DEM simulations of unsteady, homogeneous, shear flows of identical, frictional spheres under constant volume allow to investigate the fluid-solid transition. The instant application of a constant shear rate to either an isotropic, random packing (finite initial pressure) or a random, static collection of non-interacting particles (zero initial pressure) forces the system to evolve towards a steady state, while keeping constant the solid volume fraction. Some combinations of solid volume fraction and initial condition lead to a fluid-solid transition governed by the coordination number. There, the pressure does not obey the simple scaling relation observed in the solid regime \cite{sun2011}.

Section 2 describes the DEM numerical simulations performed and the specimen preparation procedure for the initial isotropic and static assembly of particles. The results of the simulations are thoroughly discussed in Section 3, while Section 4 provides some concluding remarks.

\section{DEM numerical simulations}\label{dem}

%
\begin{figure}[!h]
\centering
\includegraphics[width=4.5cm,clip]{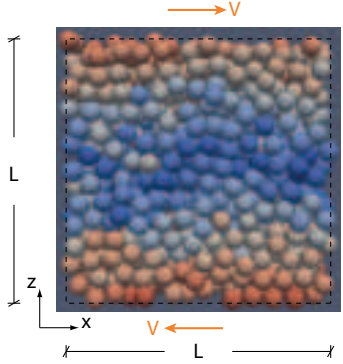}
\caption{Sketch of the flow configuration. Grey intensity indicates speed, from dark grey (blue online) -zero velocity of the particles in the core of the domain- to light grey (red online) -maximum velocity of the particles at the boundaries.}
\label{fig-1}
\end{figure}
%
\noindent We have performed DEM numerical simulations of unsteady, homogeneous shear flows of identical, frictional spheres using the open-source code Mercury-DPM \footnote{\href{www.mercurydpm.org}{www.mercurydpm.org}} \cite{tho2012,wei2012} which models the particle-particle interaction with a linear spring-dashpot contact model. Then, the normal force between particles at contact is  $\vec{f}^n_{ij}=k_n\delta_{ij}^n\vec{n}_{ij}-\gamma_n\vec{v}_{ij}^n$, where $\delta_{ij}^n$ is the normal overlap between particles, $k_n$ is the normal spring constant, $\gamma_n$ is the damping coefficient, $\vec{n}_{ij}$ is the normal unit vector and $\vec{v}_{ij}^n$ is the normal relative velocity.
Likewise, the tangential force is $\vec{f}_{ij}^t = -k_t\delta^t_{ij}\vec{t}_{ij}-\gamma_t\vec{v}_{ij}^t$,
where $k_t$ is the tangential spring constant, $\delta^t_{ij}$ the tangential overlap, $\gamma_t$ the tangential damping, $\vec{t}_{ij}$ the tangential unit vector and $\vec{v}_{ij}^t$ the tangential velocity at the contact point in the case of small overlap. The tangential overlap is set to zero at the initiation of a contact and its rate of change is given by the tangential relative velocity. The rigid body motion around the contact is taken into account to ensure that the tangential displacement always belongs to the local tangent plane of the contact \cite{lud1998,lud2008}. The magnitude of $\delta_{ij}^t$ is truncated as necessary to satisfy Coulomb law, i.e., $\left|\vec{f}_{ij}^t\right|\leq\mu\left|\vec{{f}}_{ij}^n \right|$, where $\mu$ is the inter-particle friction coefficient. The static friction coefficient is constant and equal to its dynamic counterpart. In this simplified framework, the contacts are sticking if $\left|\vec{f}_{ij}^t\right| < \mu\left|\vec{{f}}_{ij}^n \right|$ and sliding if $\left|\vec{f}_{ij}^t\right|=\mu\left|\vec{{f}}_{ij}^n\right|$.

The coefficients of normal and tangential restitution, $e_n$ and $e_t$, which relate the pre-collisional and post-collisional normal and tangential relative velocities, characterize the collisions. For the linear spring-dashpot model, the following relationships between the coefficients of restitution, the spring constants and the damping coefficients hold \cite{sil2001}:
\begin{eqnarray}
\gamma_n &=& \sqrt{\dfrac{4 m_p k_n \left(\log e_n\right)^2}{\pi^2 + \left(\log e_n\right)^2}};\nonumber\\ 
\gamma_t &=& \sqrt{\dfrac{2}{7}\dfrac{4m_{p} k_t\left(\log e_t\right)^2}{\pi^2+\left(\log e_t\right)^2}};\nonumber\\ 
k_t &=& \dfrac{2}{7}k_n\dfrac{\pi^2+\left(\log e_t\right)^2}{\pi^2+\left(\log e_n\right)^2}.
\end{eqnarray}

2000 particles of diameter $d$ and density $\rho_p$ are placed in a cubic box of dimension $L$ (\Fig\ref{fig-1}). The size $L$ of the domain depends on the desired solid volume fraction $\nu$, where, for a granular material composed of identical particles, the solid volume fraction is defined as the ratio of the particle mass density to the bulk density. In all simulations, the normal coefficient of restitution $e_n$ is equal to 0.7, the normal spring stiffness $k_n$ is equal to $10^7\mbox{ Pa}\cdot\mbox{m}$, the tangential spring stiffness $k_t$ is equal to $2/7 \ k_n$ (implying $e_t = e_n = 0.7$) and the friction coefficient $\mu$ is assumed equal to 0.3. Lees-Edwards \cite{lee1972} periodic boundary conditions along $z$ (velocity gradient direction) and periodic boundary conditions along $x$ and $y$ (flow and vorticity directions, respectively) force the system to be homogeneous in space. 
To further disregard boundary effects associated with coarse-graining \cite{wei2013}, all measurements are space-averaged over the central region, three diameters distant from the boundaries.
The relative motion of the planar boundaries perpendicular to $y$ at constant velocity $2V$ shears the system at a global shear rate $\dot\gamma = 2V/L$.

Simulations have been performed for three values of the constant solid volume fraction $\nu$: 0.59, 0.60 and 0.62. In the case of steady, homogeneous shear flows, the volume fraction determines whether the granular system is solid-like or fluid-like, with the phase transition occurring at the critical volume fraction $\nu_c$ -the largest volume fraction at which a random assembly of granular material can be sheared without developing rate-independent contributions to the stresses \cite{chi2012}.
The critical volume fraction depends on poly-dispersity \cite{oga2013,kum2014} and friction \cite{chi2012}: for identical particles with $\mu = 0.3$, $\nu_c$ is $0.596$ \cite{chi2012}. Hence, after reaching the steady state, our three specimens are in fluid ($\nu = 0.59$), solid ($\nu = 0.62$) and near-to-critical ($\nu = 0.60$) conditions.
At the steady state, there is a one-to-one relation between the critical volume fraction and the critical coordination number $Z_c$, independent of the shear rate \cite{sun2011,ves2016}. The coordination number $Z$ is an important parameter to describe the granular interaction at large volume fractions, when force chains develop \cite{ji2006}.

The evolution in time $t$ of particle pressure $p$, coordination number $Z$ and granular temperature $T$ (one third of the average square of the particle velocity fluctuations, a measure of the pseudo-thermal agitation of the particles \cite{jen1983}), is measured for different initial, isotropic, static conditions and shear rates. 

Here and in the following, the variables are made dimensionless by using the particle diameter $d$, density $\rho_p$ and normal stiffness $k_n$. Then, the dimensionless time, pressure, granular temperature and shear rate are, respectively, $t^* = t \sqrt{k_n /\left(\rho_p d^3\right)}$, $p^* = p d/k_n$, $T^* = T d\rho_p/k_n$ and $\dot\gamma^* = \dot\gamma \sqrt{\rho_p d^3/k_n}$.
%
\subsection{Preparation}\label{prepar}

The preparation of the system follows a standard three-step procedure \cite{imo2013,kum2014}:
\begin{itemize}
\item[(i)] 2000 frictionless spherical particles of diameter $d$ and density $\rho_p$ are randomly placed with random velocities in the 3D cubic box of initial size $L_0$ at moderate volume fraction $\nu_0$ (where $\nu_0 = 2000\pi d^3/6 L_0^3$), such that they have sufficient space to exchange places and randomize themselves. At the end of step (i), a static granular gas having zero pressure and coordination number is obtained. 
\item[(ii)] The friction coefficient value is imposed $\mu_0$ and the granular gas is isotropically compressed to the target volume fraction $\nu$. The cubic box reduces from size $L_0$ to $L = \left(2000\pi d^3/6\nu\right)^{1/3}$.
\item[(iii)] This is followed by a relaxation period at constant volume fraction $\nu$ to allow the particles to dissipate their kinetic energy. In order to shorten the relaxation times, an artificial background dissipation force, proportional to the velocity of the particles, is added, resembling the damping due to a background medium, such as a viscous fluid.
\end{itemize}
As was stated before, in all stages, $e_n = e_t = 0.7$ and $k_n = 10^7$ Pa$\cdot$m.
At the end of the three stages, an isotropic, static granular specimen of volume fraction $\nu$, with a certain coordination number $Z_0$ and isotropic pressure $p_0^*$ is generated.
If stages (ii) and (iii) are preformed with different values of friction $\mu_0$, different isotropic specimens (i.e, different values of $Z_0$ and $p_0^*$) are obtained.
For each volume fraction, the initial static condition can be that of an isotropic athermal (i.e., zero granular temperature) gas, with $Z_0 = 0$ and $p_0^* = 0$ or an isotropic random packing, with nonzero $Z_0$ and $p_0^*$.

%
\begin{figure}[!h]
\centering
\subfigure[\label{nu059_C0}]%
{\includegraphics[width=0.4\textwidth]{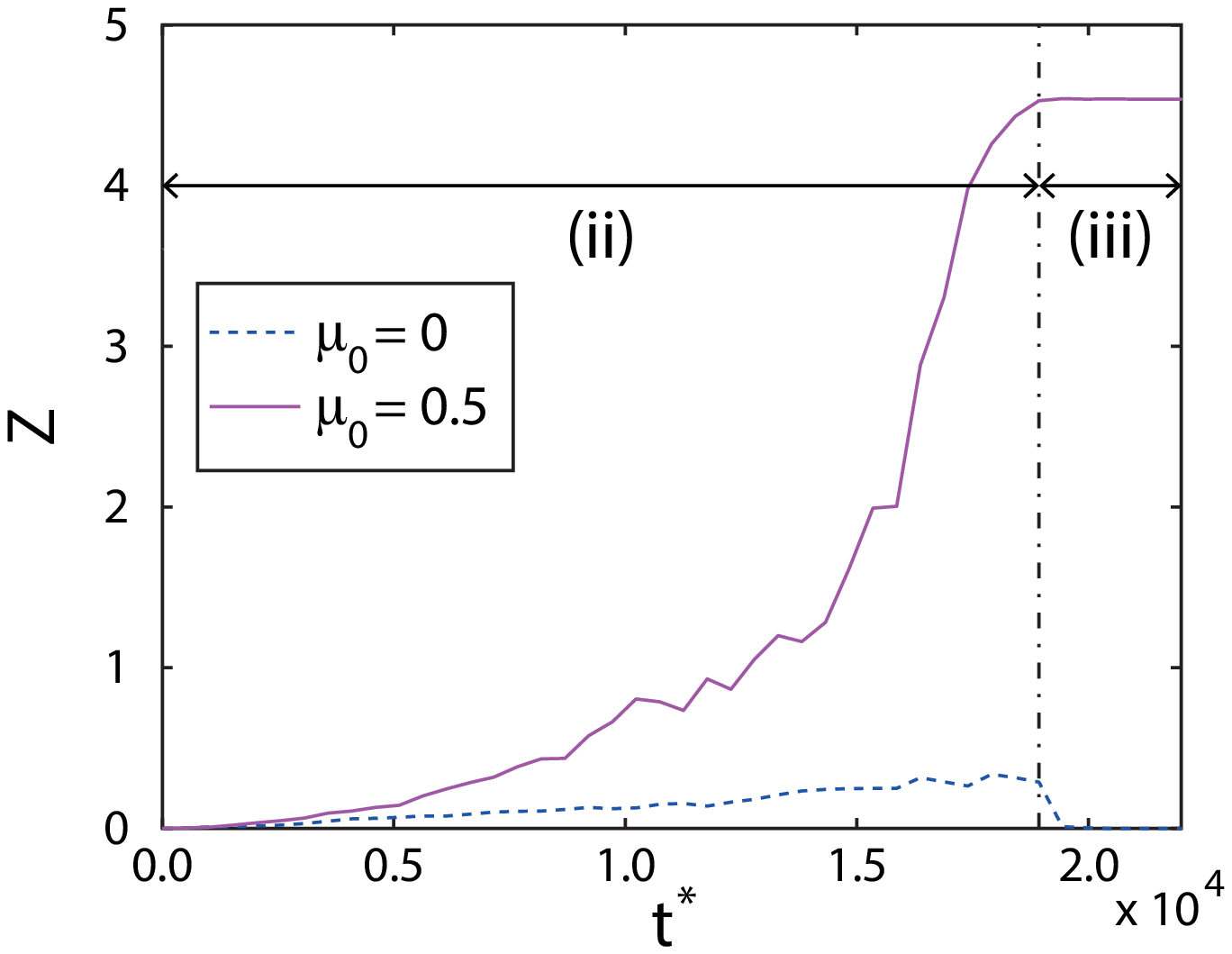}}
\subfigure[\label{nu059_p0}]%
{\includegraphics[width=0.4\textwidth]{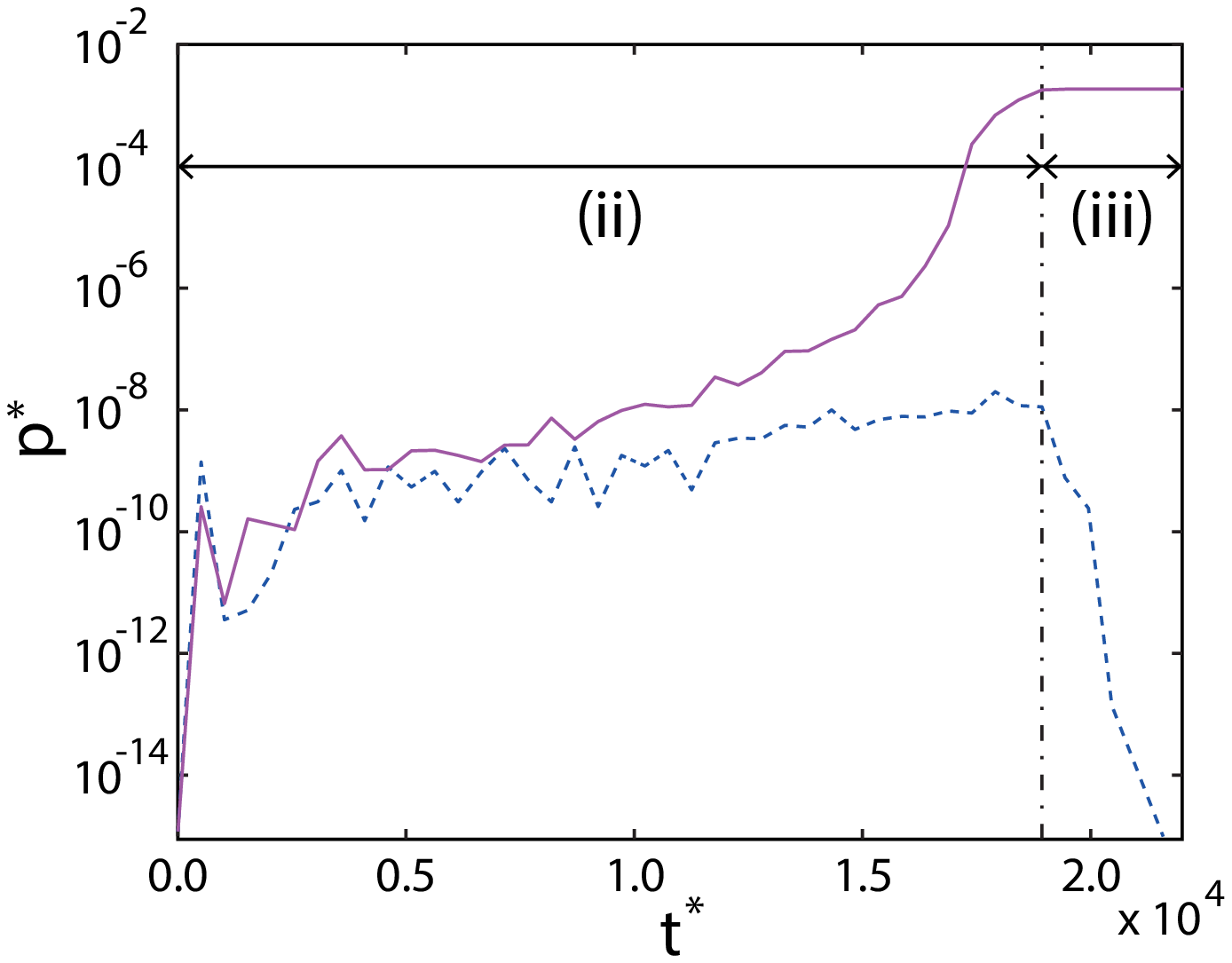}}
\caption{Time-Evolution of (a) the coordination number and (b) the dimensionless pressure during the isotropic compression (stage (ii) of the preparation procedure) and the subsequent relaxation until rest (stage (iii)) for specimens with $\nu_0=0.53$, $\nu = 0.59$ and different values of the friction $\mu_0$.}
\label{nu059_p0_C0}
\end{figure}
%
For example, \Fig\ref{nu059_p0_C0} shows the evolution of coordination number (a) and pressure (b), during stages (ii) and (iii), for $\nu_0 = 0.53$, $\nu=0.59$ and different values of $\mu_0$, respectively. At $t^* = 0$, the specimens are isotropically compressed until the target volume fraction $\nu = 0.59$ is reached ($t^* = 1.90\cdot 10^4$, stage (ii)); then, relaxation is allowed (stage (iii)). During stage (ii), both the coordination number and the pressure increase from zero to a certain value depending on the friction coefficient. When persistent, elastic contacts develop, elastic potential energy is stored.
 
If, at the end of stage (ii), the system has evolved into an isotropic random packing (for $\mu_0 = 0.5$ in \Fig\ref{nu059_p0_C0}), the coordination number and the pressure remain constant during the relaxation stage (iii) and the stored elastic energy is a signature of the force chains spanning the medium.
 
On the other hand, $Z$ and $p^*$ quickly drop to zero during stage (iii), if the system evolves towards an {i\-so\-trop\-ic} athermal gas (case $\mu_0 = 0$  in \Fig\ref{nu059_p0_C0}). In this case, there are no persistent contacts among the grains and the potential energy turns into kinetic energy, which is finally dissipated through collisions at the end of stage (iii). Similar behaviours are obtained for other values of $\nu$.

Parameters $\nu_0$ and $\mu_0$ adopted in the preparation stages are summarized in \Tab\ref{tab1}(a) and (b), with the measured values of the coordination number $Z_0$ and dimensionless pressure $p_0^*$. The values of $\mu_0$ were chosen to obtain isotropic random packings at different volume fractions with roughly the same pressure $p_0^* \approx 2 \cdot 10^{-3}$.

At the end of the preparation procedure, the friction $\mu$ is set to 0.3, and the velocities at the boundaries are switched on so that the global shear rate $\dot\gamma$ has the desired value, constant in time.

\begin{table}[!h]
\centering
\caption{Summary of the parameters used in the three-step preparation procedure and corresponding measured values of the coordination number and the dimensionless pressure before the beginning of the homogeneous shearing.}
\label{tab1}
\subtable[Isotropic athermal gas]{%
\begin{tabular}{ccccc}
\hline\noalign{\smallskip}
$\nu$ & $\nu_0$ & $\mu_0$ & $p_0^*$ & $Z_0$  \\
\noalign{\smallskip}\hline\noalign{\smallskip}
0.59 & 0.53 & 0 & 0 & 0\\
0.60 & 0.56 & 0 & 0 & 0\\
0.62 & 0.58 & 0 & 0 & 0\\
\noalign{\smallskip}\hline
\end{tabular}
}\\
\subtable[Isotropic random packing]{%
\label{tab2}
\begin{tabular}{ccccc}
\noalign{\smallskip}\hline
$\nu$ & $\nu_0$ & $\mu_0$ & $p_0^*$ & $Z_0$  \\
\noalign{\smallskip}\hline\noalign{\smallskip}
0.59 & 0.53 & 0.5 & $1.85 \cdot 10^{-3}$ & 4.54\\
0.60 & 0.56 & 0.5 & $2.05 \cdot 10^{-3}$ & 4.73\\
0.62 & 0.58 & 0.1 & $2.12 \cdot 10^{-3}$ & 5.75\\
\noalign{\smallskip}\hline
\end{tabular}
}
\end{table}
%

\section{Results}\label{results}

This Section describes the results of the DEM simulations.
%
First, in order to define the mechanical quantities governing the phase transition in granular materials under both static and dynamic conditions (the latter being either steady and unsteady), the results of the preparation procedure on specimens at different $\mu_0$ and $\nu$ are used to draw a phase diagram in terms of inter-particle friction, volume fraction and coordination number. \\
Then, the fluid-solid transition in the unsteady regime is analysed in terms of evolution of coordination number, pressure and granular temperature with the accumulated shear strain $\gamma = \dot\gamma t$. 
In particular, the results obtained from slowly shearing initially isotropic athermal gases emphasize the role of the solid volume fraction. Those obtained considering initially isotropic random packings allow to appreciate the role of the initial conditions of the assembly, and, finally, different shear rates at sufficiently large values of volume fraction are accounted for to show the rate-independency attained once the granular material solidifies.

\subsection{Phase diagram}\label{jamm}

\noindent The fluid-solid transition in granular materials has been studied in the literature in both static and dynamic conditions, with different protocols \cite{liu1998,ohe2002,maj2007,son2008,hat2008,ots2009,sil2010,zha2011,sun2011,chi2012}.
 
In isotropic, static conditions, the transition from a zero pressure state to a nonzero pressure state is known as jamming transition \cite{liu1998,son2008,sil2010}. Isotropic granular packings, characterized by nonzero pressure and coordination number, are jammed structures, with a contact network percolating in all directions. The jamming transition is defined as a mechanical stable state characterized by zero pressure and finite coordination number, occurring when the system approaches a critical solid volume fraction $\nu_J$ (minimum volume fraction at which a random isotropic packing exists), whose value depends on the preparation history of the packing \cite{sil2010}. Moreover, there is a relation between $\nu_J$ and the coordination number at the jamming transition, $Z_J$.
The effect of friction on the transition has been numerically investigated by approaching the jamming from below (i.e., isotropically compressing an athermal granular gas \cite{son2008}) and from above (i.e., isotropically decompressing a random packing \cite{sil2010}). 
The results of the two procedures qualitatively agree and show that both the volume fraction and the coordination number at the jamming transition monotonically decreases for increasing friction. 

In the case of steady, homogeneous shearing flows, Chialvo et al. \cite{chi2012} have measured the critical volume fraction $\nu_c$ at which rate-independent components of the stresses develop, based on the observation that the pressure fluctuations peak at that transition. Sun and Sundaresan \cite{sun2011} identified the critical coordination number $Z_c$ governing the same transition by extrapolation to zero of the relation between the pressure and the coordination number. Simple fitting expressions \cite{sun2011} provide the dependence of both $\nu_c$ and $Z_c$ on the friction coefficient:
\begin{eqnarray}
\nu_c &=& 0.582 + 0.058 \exp(-5\mu);\label{fitnuc}\\
Z_c &= &4.15 + 1.85 \exp(-5\mu).\label{fitCc}
\end{eqnarray}

\begin{figure}[!h]
\centering
\includegraphics[width=0.4\textwidth]{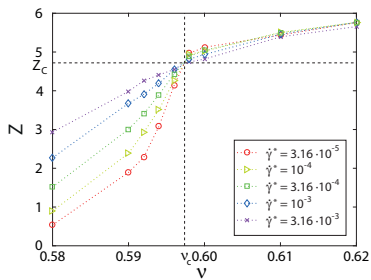}
\caption{Coordination number versus solid volume fraction in steady, homogeneous shear flows with $\mu=0.3$ and different values of the dimensionless shear rate.}
\label{CoordNumber}
\end{figure}

In the same case of steady, homogeneous shearing flows, Ji and Shen (for frictional spheres with $\mu = 0.5$ \cite{ji2006}) and Vescovi and Luding (for frictionless spheres \cite{ves2016}) have studied the dependence of the coordination number on the volume fraction, for different values of the dimensionless shear rate $\dot\gamma^*$ (or, equivalently, the dimensionless particle stiffness). 
They found that the coordination number increases with increasing solid volume fraction. Curves at different $\dot\gamma^*$ intersect at a critical point $\nu_c$-$Z_c$, which is in excellent agreement with the fitting expressions of \Eqs\eqref{fitnuc} and \eqref{fitCc}. 
\Fig\ref{CoordNumber} shows the coordination number as a function of the solid volume fraction obtained in the present simulations of steady, homogeneous shear flows for different values of the scaled shear rate when $\mu=0.3$. The steady values of $Z$ have been obtained by averaging in time over at least 1000 time steps. All the curves intersect at $\nu_c \approx 0.596$, with $Z_c \approx 4.6$. Once again, these values are in good agreement with \Eqs\eqref{fitnuc} and \eqref{fitCc}.
%
%

%
\begin{figure}[!h]
\centering
\subfigure[\label{mu_nuc}]%
{\includegraphics[width=0.4\textwidth]{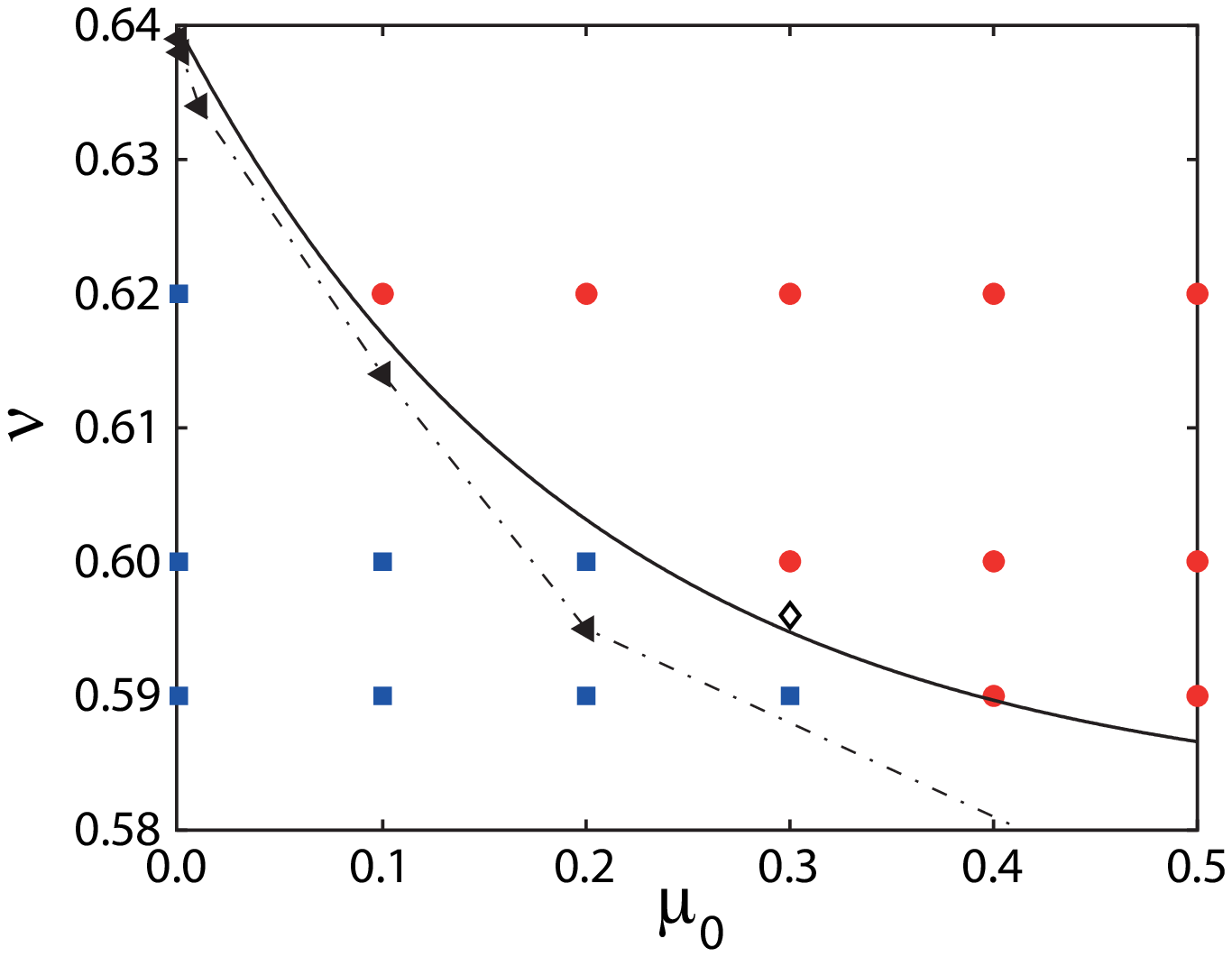}}
\subfigure[\label{mu_Cc}]%
{\includegraphics[width=0.4\textwidth]{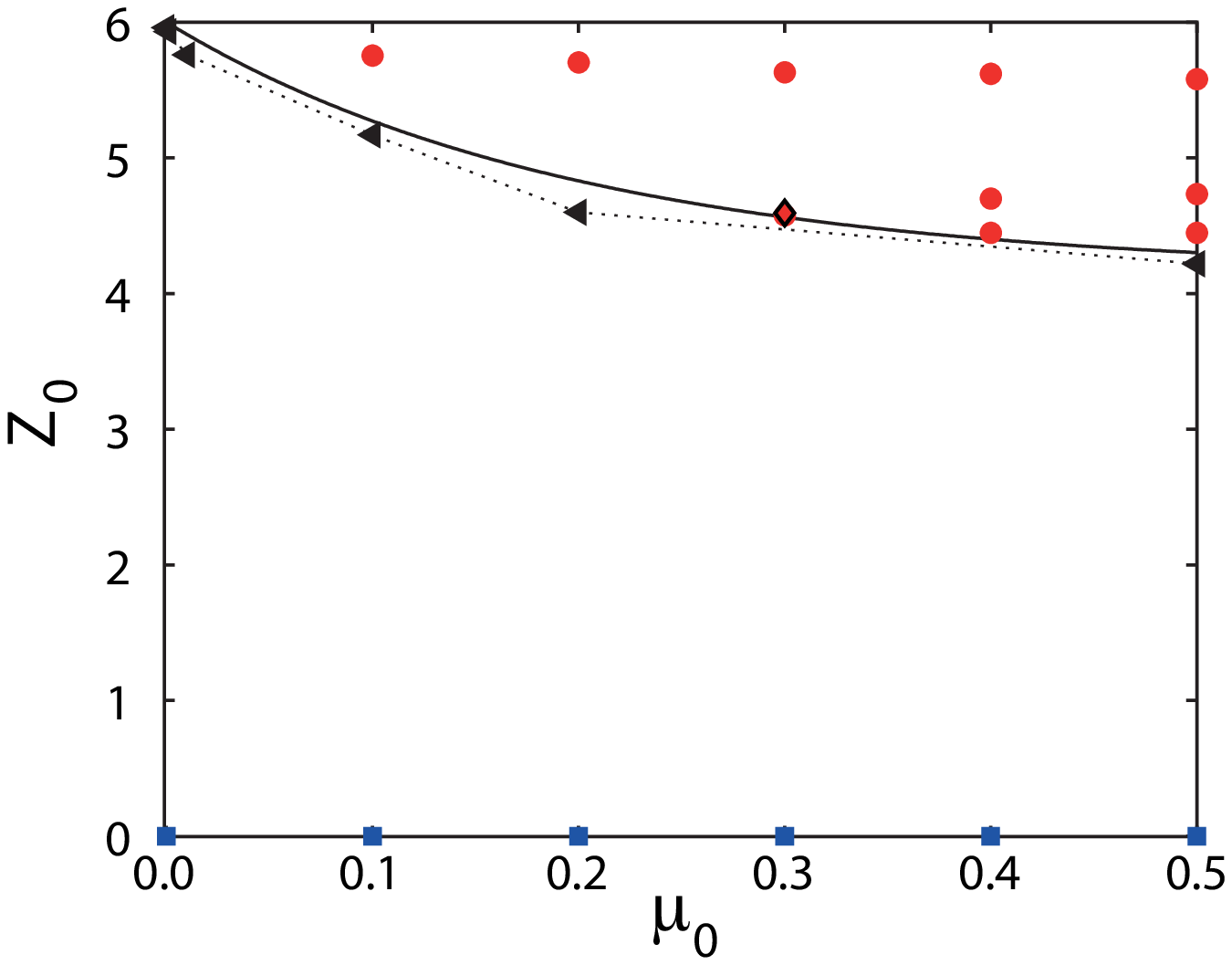}}%
\caption{Phase diagrams of the initial conditions obtained at the end of the three-step preparation procedure in terms of (a) volume fraction  and (b) coordination number versus the friction coefficient. The initial, static specimen is an isotropic, athermal granular gas (zero coordination number, squares) or an isotropic random packing (finite coordination number, circles). Also plotted are the numerical results for the jamming transition of isotropic random packings (triangles \cite{sil2010}) and the critical values $\nu_c = 0.596$ and $Z_c = 4.6$ inferred from \Fig\ref{CoordNumber} (diamonds). The solid lines in \Fig\ref{mu_nuc_Cc}(a) and (b) represent \Eqs\eqref{fitnuc} and \eqref{fitCc}, describing the phase transition in steady, simple shearing.}
\label{mu_nuc_Cc}
\end{figure}
%
\Fig\ref{mu_nuc_Cc} depicts the phase diagrams obtained at the end of stage (iii) of our preparation procedure, in terms of solid volume fraction $\nu$ and coordination number $Z$ versus friction coefficient $\mu_0$. The boundary between isotropic athermal gases and isotropic random packings obtained with the present protocol is, perhaps surprisingly, well identified by \Eq\eqref{fitnuc}, which, however, do not describe the jamming transition for isotropic random packings observed by Silbert \cite{sil2010} (\Fig\ref{mu_nuc}). As demonstrated in \cite{son2008}, the volume fraction at the jamming point is not unique, but depends on the preparation protocol.

On the other hand, when plotted in terms of coordination number versus friction coefficient (\Fig\ref{mu_Cc}), \Eq\eqref{fitCc} correctly represents the lower limit for the existence of isotropic random packings obtained with the present preparation procedure, and also fits the jamming transition in isotropic random packings observed in \cite{sil2010} and the critical coordination number of steady, homogeneous shear flows of \Fig\ref{CoordNumber}. 

\subsection{Unsteady regime}

\subsubsection{Slow shearing of initially isotropic athermal gases}\label{slow}

\begin{figure}[!h]
\centering
\subfigure[\label{fl_gp01_C}]%
{\includegraphics[width=0.4\textwidth]{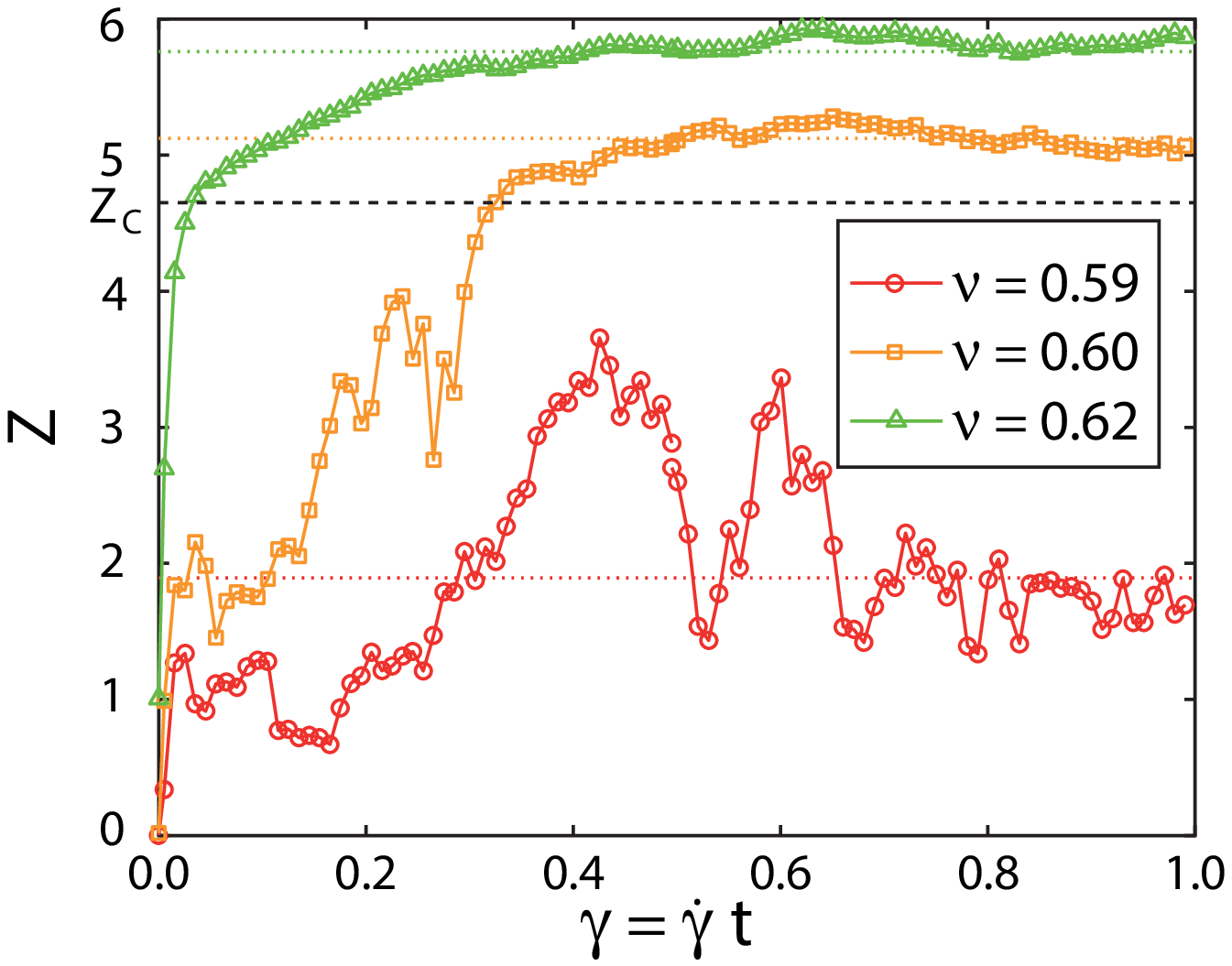}}
\subfigure[\label{fl_gp01_p}]%
{\includegraphics[width=0.4\textwidth]{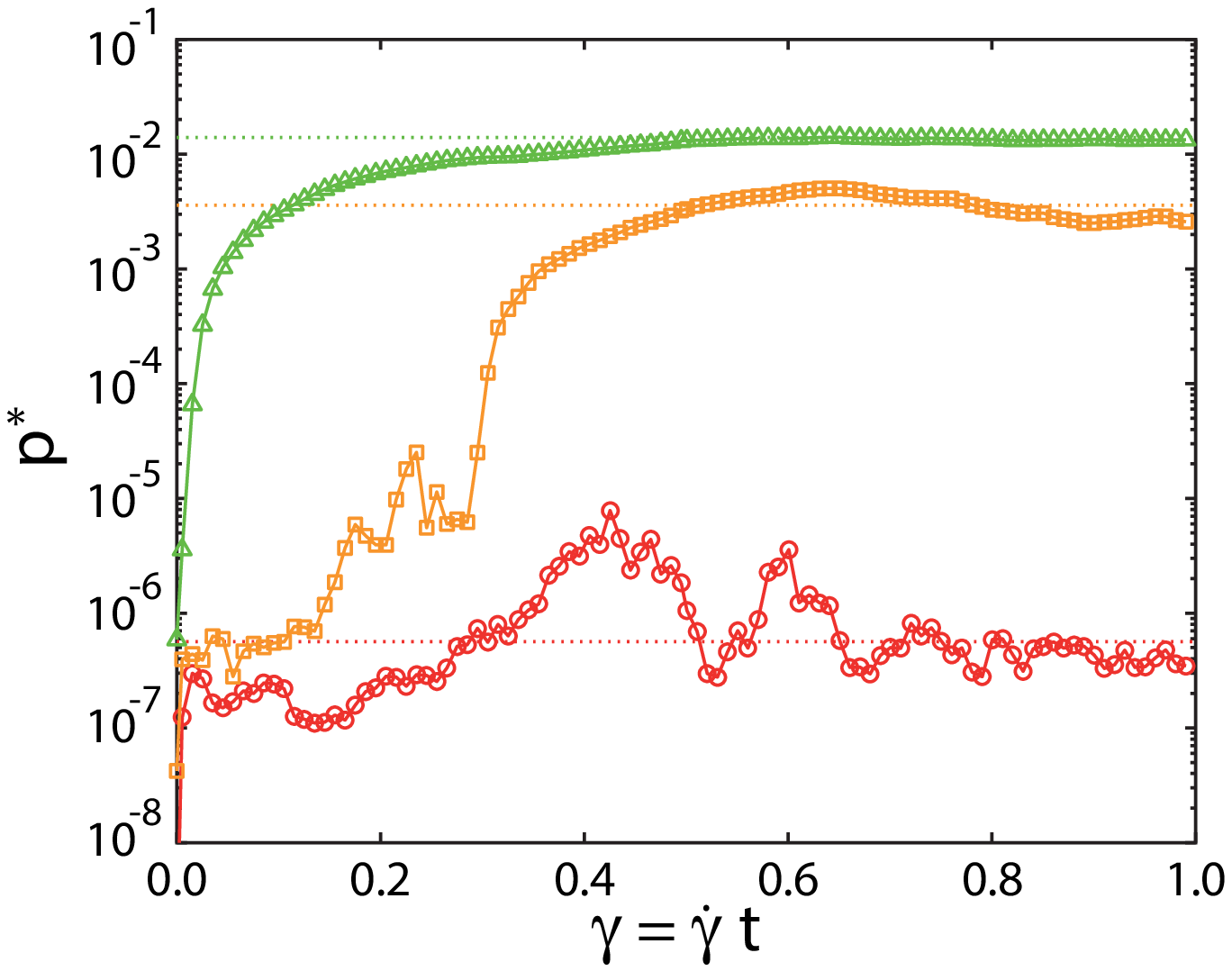}}
\caption{Evolution of (a) the coordination number and (b) the dimensionless pressure for different values of the volume fraction when $\dot\gamma^* = 3.16 \cdot 10^{-5}$ and the specimen is, initially, an isotropic athermal gas ($p_0^* =0$, $Z = 0$). Final steady state values are denoted with dotted lines; the dashed line represents the critical coordination number of \Eq\eqref{fitCc}.}
\label{fl_gp01}
\end{figure}

\noindent The evolutions of the coordination number $Z$ and the dimensionless pressure $p^*$ with the accumulated shear strain $\gamma$ obtained by slowly shearing initially isotropic athermal gases are illustrated in \Fig\ref{fl_gp01}(a) and (b), respectively. Here, and in what follows, each point represents the average of 10 measurements over a window of amplitude $\Delta\gamma = 10^{-2}$. 

At $t = 0$ ($\gamma = 0$), a small, dimensionless shear rate $\dot\gamma^*= 3.16 \cdot 10^{-5}$ is instantly applied. This can be equivalently interpreted as imposing a small dimensional shear rate $\dot\gamma$ or a large particle stiffness $k_n$. During the transient regime, the coordination number and the pressure are homogeneous along the flow depth (i.e., constant in space) at each time step, whereas the velocity profiles are initially not linear (for $\gamma < 0.05$) as, instead, in simple shear flows. This is due to the instantaneous application of a relative velocity between the boundaries which does not immediately affect the entire granular material. Nevertheless, the non-homogeneity disappears for $\gamma > 0.05$, where $\dot\gamma^*$ is homogeneous in space and constant in time.

%
%
%
\begin{figure}[!h]
\centering
\subfigure[\label{fl_gp01_stdC}]%
{\includegraphics[width=0.4\textwidth]{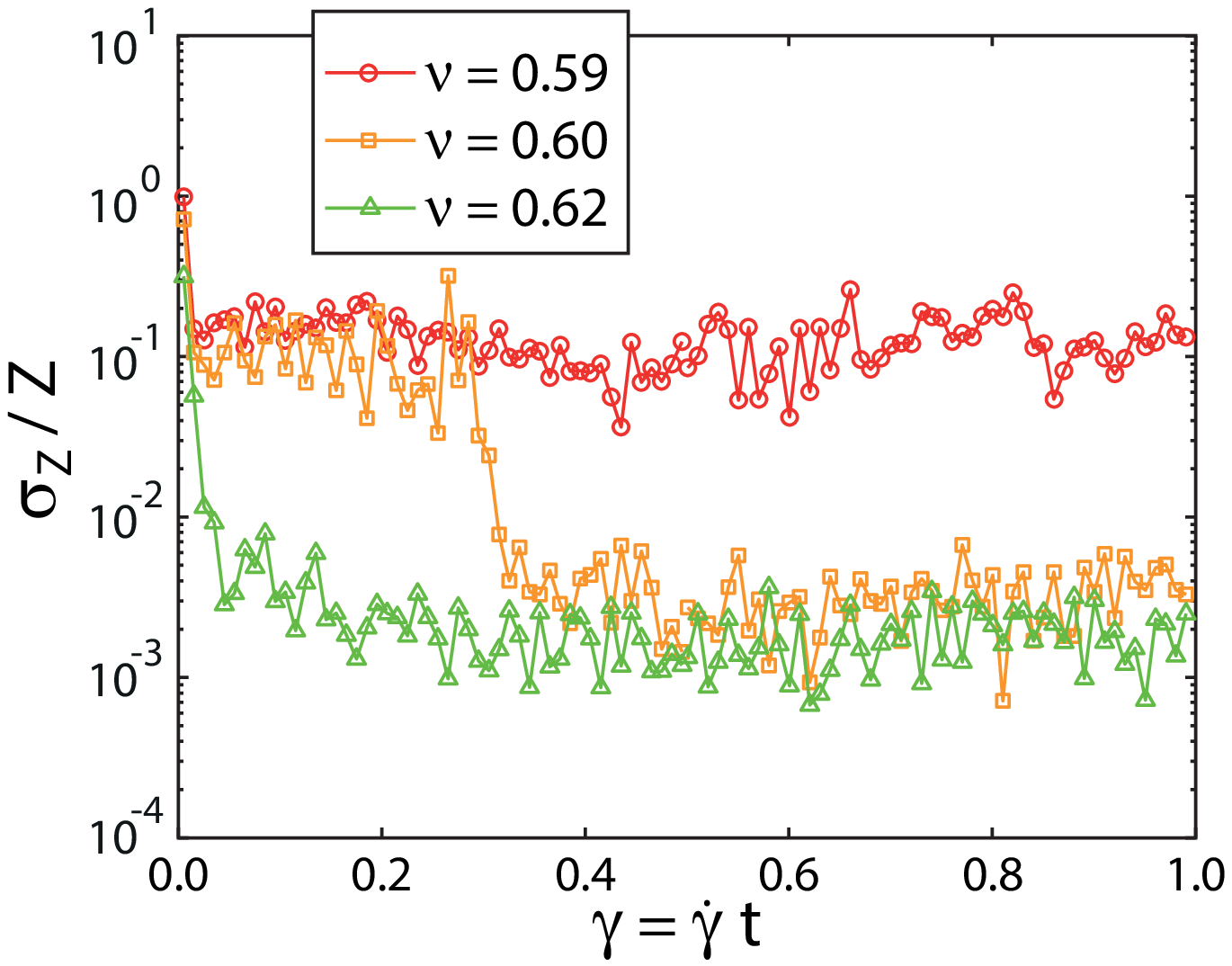}}
\subfigure[\label{fl_gp01_stdp}]%
{\includegraphics[width=0.4\textwidth]{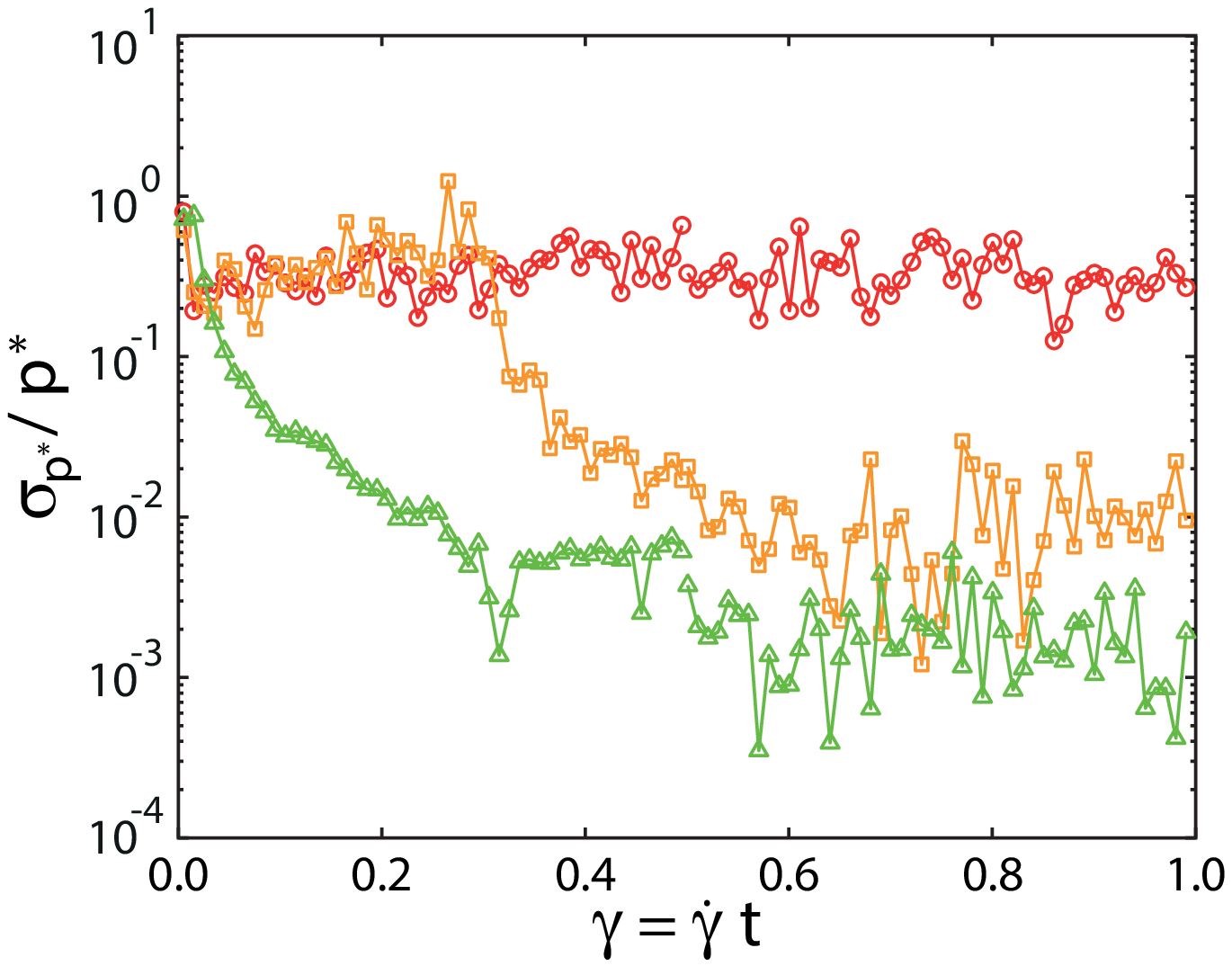}}
\caption{Evolution of the scaled standard deviation of (a) the coordination number and (b) the dimensionless pressure for different values of the volume fraction when $\dot\gamma^* = 3.16 \cdot 10^{-5}$ and the specimen is, initially, an isotropic athermal gas ($p_0^* =0$, $Z = 0$).}
\label{fl_gp01_std}
\end{figure}
%

\Fig\ref{fl_gp01} shows that the coordination number and the pressure essentially increase with time until the steady state is reached, and are closely related, as was previously observed by Sun and Sundaresan~\cite{sun2011} in the solid regime. 

In the case $\nu = 0.59$, the steady state is rapidly reached, with large fluctuations for both $p^*$ and $Z$. This is a signal that the system is always fluid during the simulation, with large fluctuations in the coordination number due to the rapid rearrangement of the particles. The coordination number is always less than 4.6 (dashed lines in \Fig\ref{fl_gp01}b), i.e., the critical coordination number for the development of rate-independent component of the stresses in steady, homogeneous shear flows (\Fig\ref{CoordNumber}), but also greater than zero, indicating the presence of particle clusters. The continuous destruction and re-building of clusters gives rise to the observed large fluctuations in both coordination number and pressure. To characterize the fluctuations, in \Fig\ref{fl_gp01_std}(a) and (b) the standard deviations of the coordination number $\sigma_Z$ and pressure $\sigma_{p^*}$ over the window $\Delta\gamma$, scaled by the corresponding quantities $Z$ and $p^*$, as functions of $\gamma$ are plotted. 
When $\nu=0.59$, the scaled amplitude of the fluctuations in the coordination number is of order $10^{-1}$ (\Fig\ref{fl_gp01_std}(a)).

%
For $\nu = 0.62$, the system presents small fluctuations in both $p^*$ and $Z$ during the whole process (\Fig\ref{fl_gp01}). Indeed, $\sigma_{Z}/Z$ and $\sigma_{p^*}/p^*$ are of order of $10^{-3}$  after a few time steps (\Fig\ref{fl_gp01_std}): this is interpreted as a signature of solid behaviour as long as the coordination number is larger than 4.6. 
The standard deviation of each quantity is affected by the number of measurements within $\Delta \gamma$ and the amplitude of $\Delta \gamma$. Nevertheless, $\sigma_{Z}/Z$ and $\sigma_{p^*}/p^*$ for $\nu = 0.62$ are always two orders of magnitude smaller than in the case $\nu=0.59$, independently of both $\Delta \gamma$ and the number of measurements.

The coordination number and the pressure smoothly increase from $0$ to the steady values, with the latter one larger than that at $\nu=0.59$ by several orders of magnitude. The steady state is reached at $\gamma \approx 0.5$. The chains of contacts spanning the entire domain require time to develop within the specimen and, as a consequence, a clear transient regime is shown in terms of both $Z$ and $p^*$. Once the contact network develops, the very small fluctuations of $Z$ and $p^*$ are due to micro-structural rearrangements during shear. The larger the volume fraction, the smaller the fluctuations, since particles are more compacted and cannot easily abandon force chains.
We point out that, even in solid conditions, no shear localization is observed and the specimens are homogeneous during the whole transient regime. 

Finally, the transition from fluid to solid is evident in the case $\nu = 0.60$. There, the behaviour of $Z$ and $p^*$ is initially very similar to the fluid case ($\nu=0.59$), with large fluctuations in both quantities whose scale amplitude is of order $10^{-1}$ (\Fig\ref{fl_gp01_std}). At $\gamma \approx 0.3$, the system experiences a significant increase in the coordination number and pressure (\Fig\ref{fl_gp01}), with a corresponding decrease in the scaled amplitude of the fluctuations, now of the order of $10^{-3}$(\Fig\ref{fl_gp01_std}). At $\gamma \approx 0.3$, the coordination number is approximately equal to 4.6, that is the critical value $Z_C$ of \Eq\eqref{fitCc}.

This analysis suggests that the critical coordination number for the development of rate-independent components of the stresses evaluated in steady, homogeneous shear flows, and for the jamming transition of isotropic random packings, coincides with that evaluated in case of unsteady, homogeneous shear flows.

%
\begin{figure}[!h]
\centering
\includegraphics[width=0.4\textwidth]{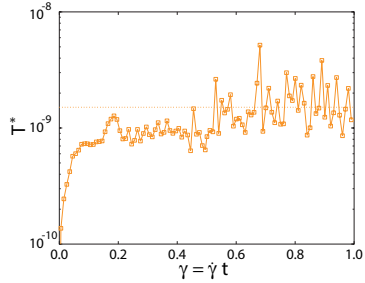}
\caption{Evolution of dimensionless granular temperature $T^*$ for $\nu = 0.60$ when $\dot\gamma^* = 3.16 \cdot 10^{-5}$ and the specimen is, initially, an isotropic athermal gas ($p_0^* =0$, $Z = 0$). The dotted line represents the value at steady state.}
\label{fl_gp01_T}
\end{figure}
%
\Fig\ref{fl_gp01_T} illustrates the evolution of the dimensionless granular temperature when $\nu = 0.60$. Unlike the pressure and the coordination number, the fluctuations of $T^*$ are small when $Z<Z_C$ and large when $Z>Z_C$. When the granular material solidifies ($Z>Z_C$), there are force chains spanning the domain that are continuously broken and re-formed during the shear. A cascade of collisions is generated by the breaking of the force chains, resulting in strong fluctuations of granular temperature. This slightly affects the pressure because, for sufficiently rigid particles, the elastic component of the pressure is much larger than that associated with exchange of momentum \cite{ber2015b}. In absence of a contact network spanning the whole medium (under fluid conditions, i.e., when $Z<Z_C$), the fluctuations in the granular temperature are far less dramatic (\Fig\ref{fl_gp01_T}), since the flow is agitated but continuum without abrupt events.

\subsubsection{Slow shearing of initially isotropic random packings}\label{init_cond}
%
\begin{figure}[!h]
\centering
\subfigure[\label{sol_gp01_C}]%
{\includegraphics[width=0.4\textwidth]{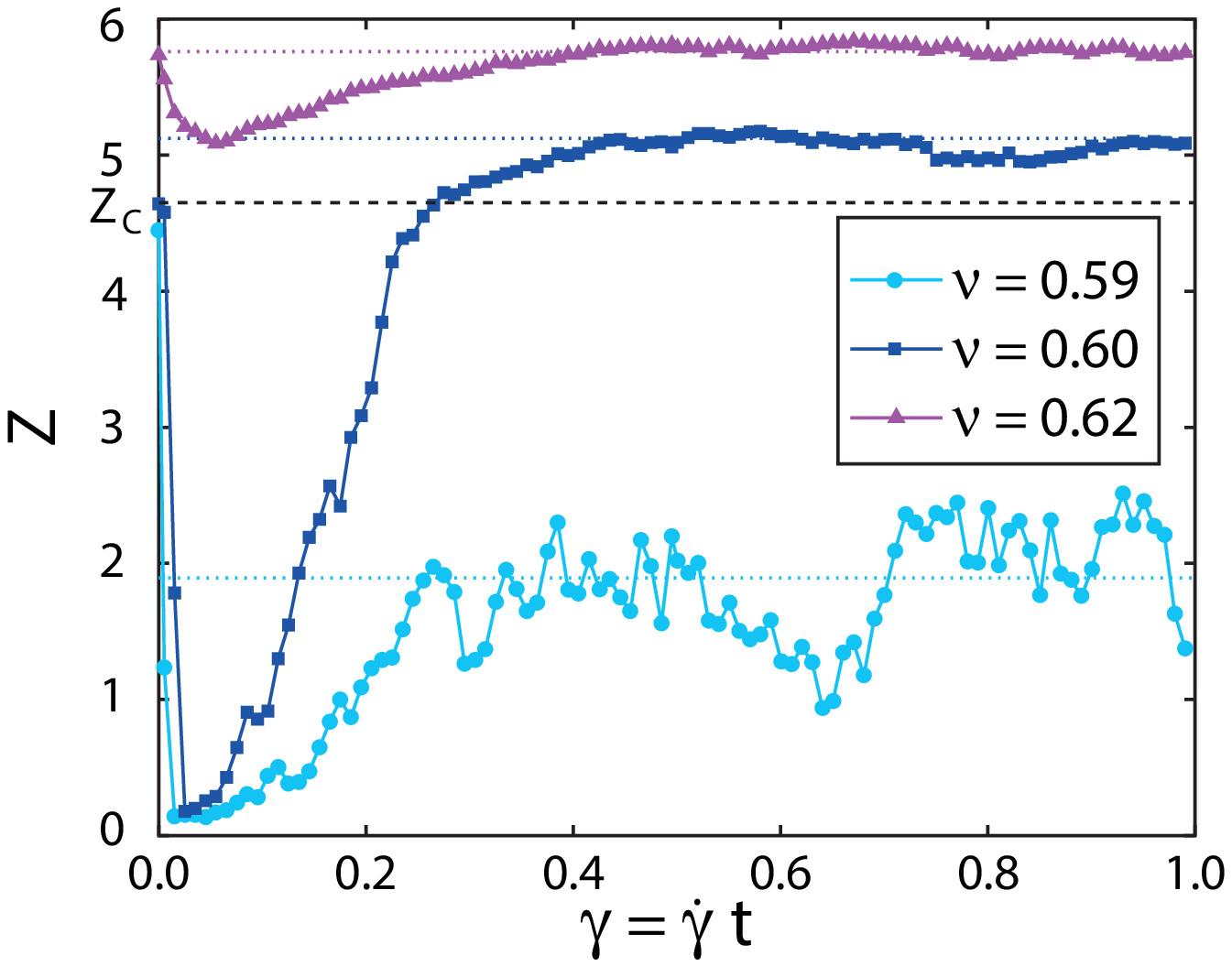}}
\subfigure[\label{sol_gp01_p}]%
{\includegraphics[width=0.4\textwidth]{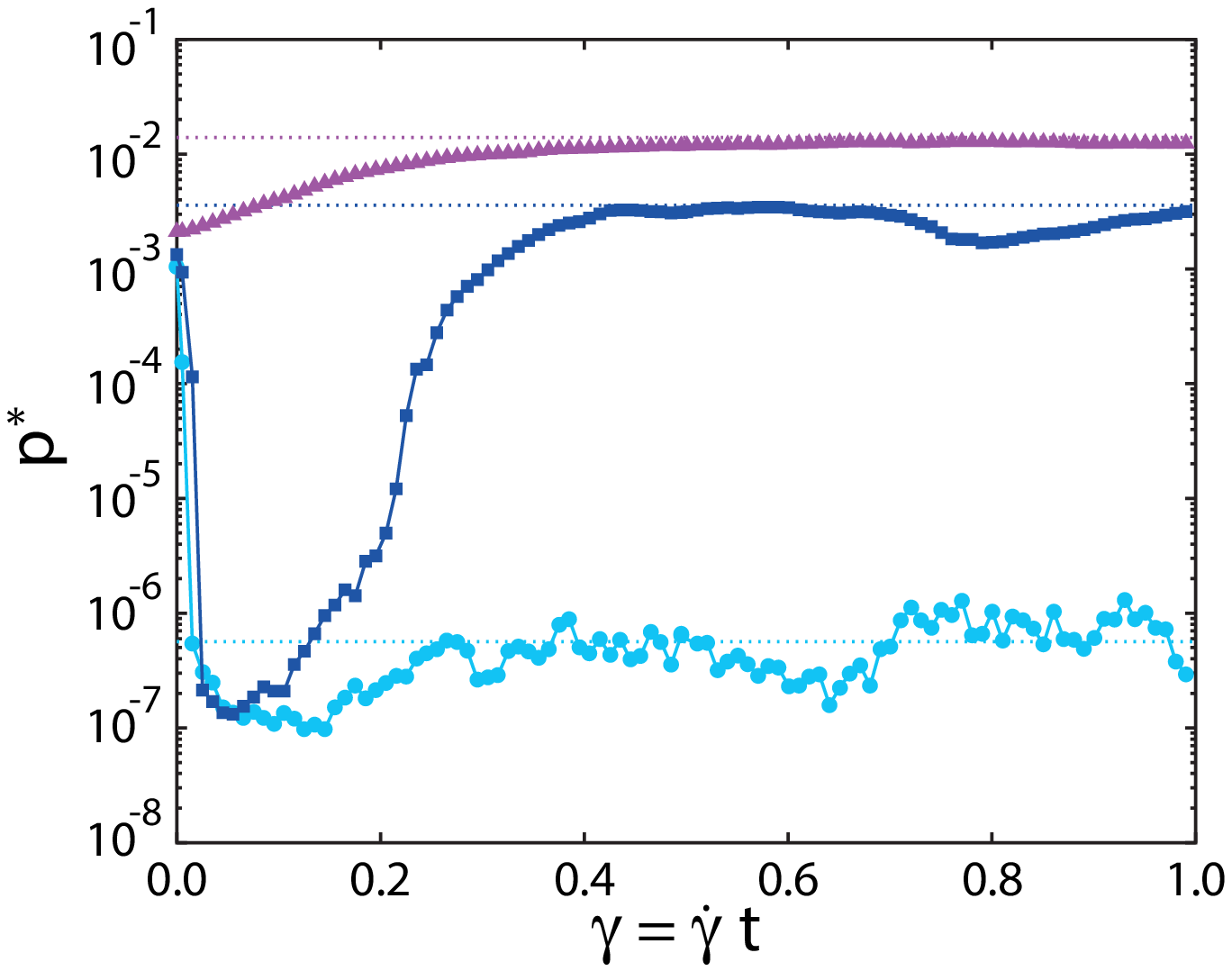}}
\caption{Evolution of (a) the coordination number and (b) the dimensionless pressure for different values of the volume fraction when $\dot\gamma^* = 3.16 \cdot 10^{-5}$ and the specimen is, initially, an isotropic random packing, with $p_0^* \approx 2\cdot 10^{-3}$. Steady state values are denoted with dotted lines; the dashed line represents the critical coordination number of \Eq\eqref{fitCc}.}
\label{sol_gp01}
\end{figure}
%
%
\begin{figure}[!h]
\centering
\subfigure[\label{sol_gp01_stdC}]%
{\includegraphics[width=0.4\textwidth]{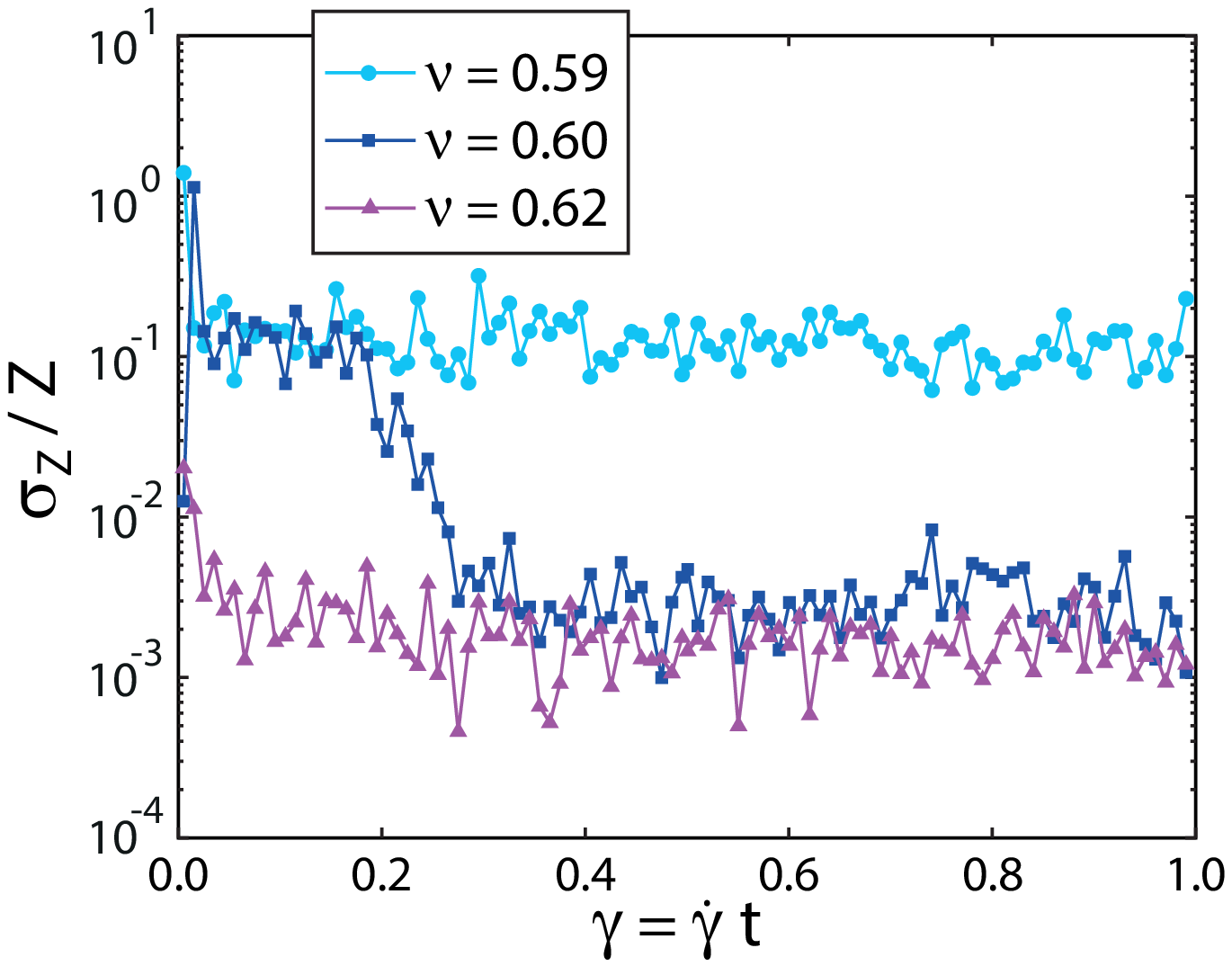}}
\subfigure[\label{sol_gp01_stdp}]%
{\includegraphics[width=0.4\textwidth]{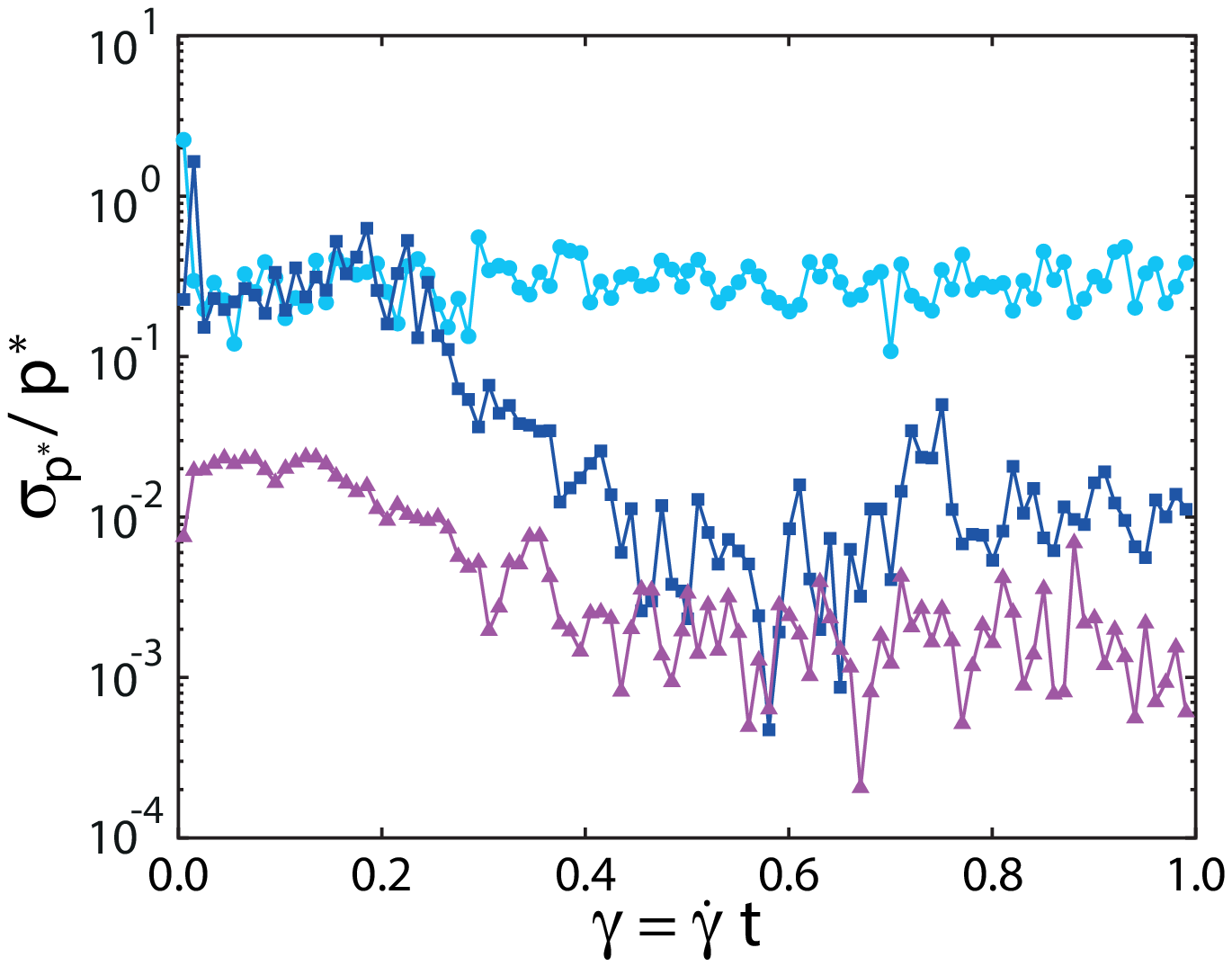}}
\caption{Evolution of the scaled standard deviations of (a) the coordination number and (b) the dimensionless pressure for different values of the volume fraction when $\dot\gamma^* = 3.16 \cdot 10^{-5}$ and the specimen is, initially, an isotropic random packing, with $p_0^* \approx 2\cdot 10^{-3}$. }
\label{sol_gp01_std}
\end{figure}
\noindent \Fig\ref{sol_gp01} depicts the evolution of $Z$ and $p^*$ when initially isotropic random packings, with initial dimensionless pressure $p^*_0 \approx 2\cdot 10^{-3}$, are sheared by imposing $\dot\gamma^* = 3.16 \cdot 10^{-5}$. The corresponding standard deviations are reported in \Fig\ref{sol_gp01_std}.

In the cases $\nu = 0.59$ and $\nu = 0.60$, the coordination number rapidly drops to zero during the first time steps (\Fig\ref{sol_gp01}a). The instantaneous application of the shearing breaks all the force chains within the initial packing and the system fluidizes ($Z\approx 0$). Then, $Z$ and $p^*$ increase, mimicking \Fig\ref{fl_gp01}. As for the case of slow shearing of initially athermal gases, the scaled amplitudes of the fluctuations of coordination number and pressure is orders of magnitude larger if the system is in the fluid regime ($Z<Z_C$). 

The response of the material is very different at $\nu = 0.62$ with respect to the case of an initially isotropic athermal gas. If the system is initially an isotropic random packing, the pressure monotonically increases from the static to the steady value (\Fig\ref{sol_gp01}b), whereas the coordination number initially decreases ($\gamma < 0.06$) and then increases (\Fig\ref{sol_gp01}a), while remaining above $Z_C$. The solid behaviour is confirmed by the small fluctuations of both $Z$ and $p^*$ (\Fig\ref{sol_gp01_std}). The non monotonic relation between the pressure and the coordination number can be understood given that the system must evolve from an isotropic to an anisotropic solid state. At the beginning, all contacts carry the same load. We can imagine that, during the initial stage, some contacts are lost due to shear, while the remaining experience an increase in their loads. As a consequence, the coordination number, which includes particles with zero (floaters) or one contact (rattlers) \cite{gon2009,kum2014}, even if they do not participate to the contact network, decreases even though the pressure increases.\\
The phenomenon numerically observed in case of $\nu = 0.59$ and $0.60$ (\Fig\ref{sol_gp01}a and b) is analogous to what experimentally obtained in case of loose sands and known as static liquefaction in the geotechnical literature \cite{cas1969,lad1992,dip1995}.

\subsubsection{Dependence on the shear rate}\label{shear_rate}

\noindent So far, it has been suggested that granular materials subjected to unsteady, homogeneous shearing experience a fluid-solid transition when the coordination number exceeds a critical value; and the transition has been identified by observing completely different amplitudes in the fluctuations of pressure and coordination number, at least for slow shearing and/or sufficiently rigid particles, in the two regimes. Although a criterion based on the fluctuations in the pressure has been already suggested for identifying a phase transition in the case of steady, homogeneous shearing of granular materials \cite{chi2012}, it has also been shown that phase transition implies the development of {rate-in\-de\-pend\-ent} components of stresses. 

%
\begin{figure}[!h]
\centering
\subfigure[\label{nu059_C}]%
{\includegraphics[width=0.4\textwidth]{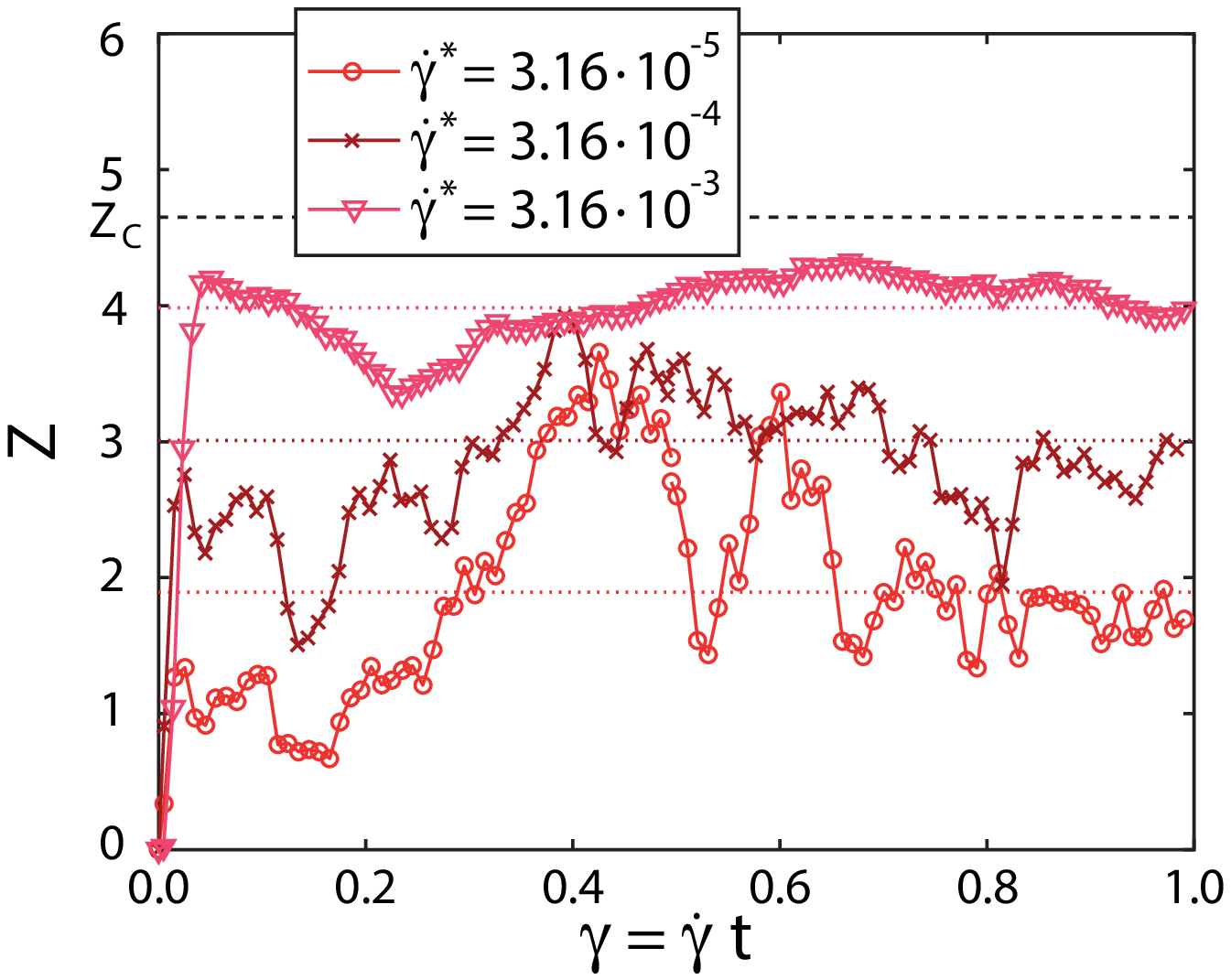}}
\subfigure[\label{nu059_stdC}]%
{\includegraphics[width=0.4\textwidth]{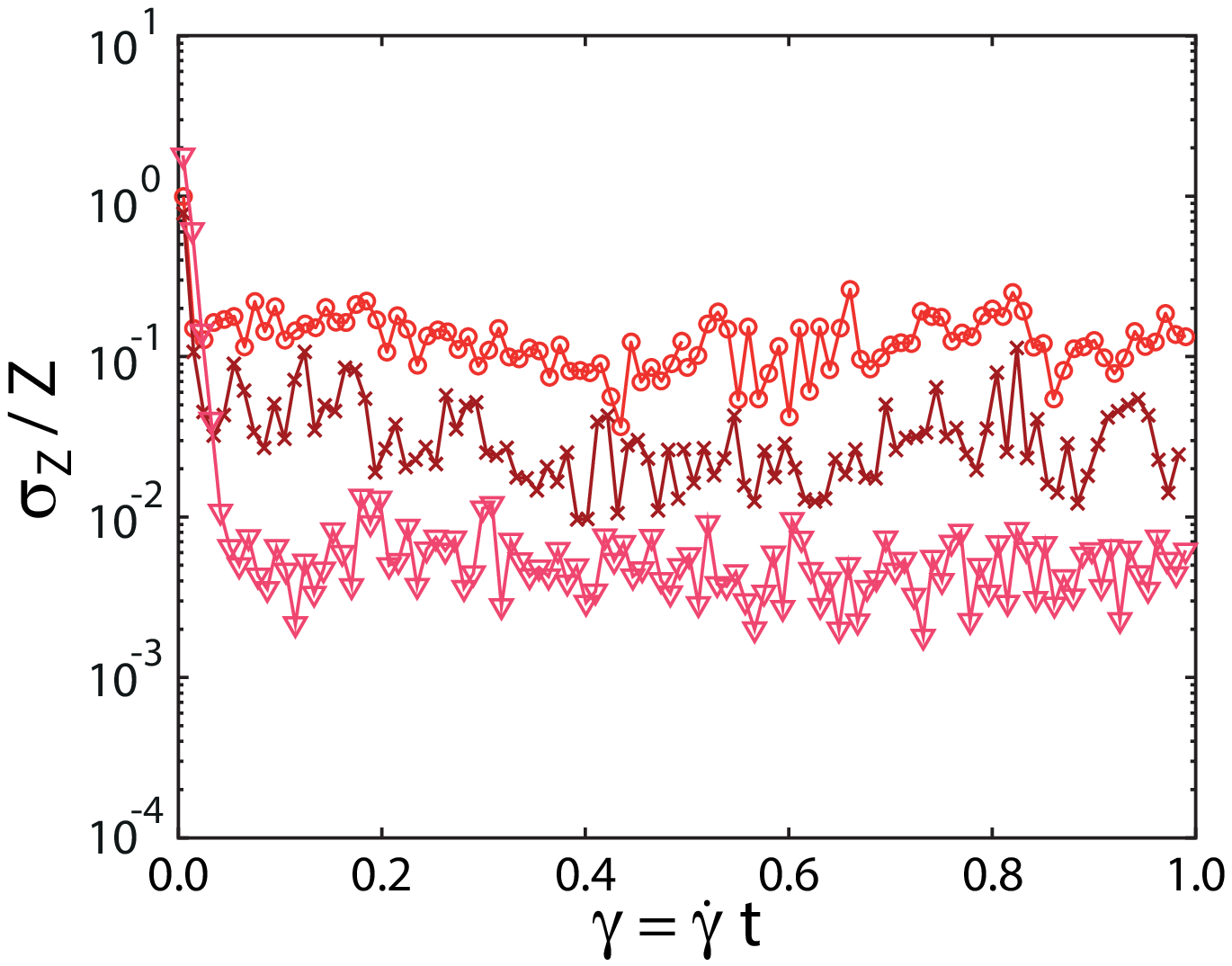}}
\caption{Evolution of (a) the coordination number and (b) the scaled standard deviation of the coordination number for different values of the dimensionless shear rate, when $\nu = 0.59$ and the specimen is, initially, an isotropic athermal gas. Steady state values are denoted with dotted lines; the dashed line represents the critical coordination number.}
\label{nu059_CstdC}
\end{figure}
%
%
\begin{figure}[!h]
\centering
\subfigure[\label{nu060_C}]%
{\includegraphics[width=0.4\textwidth]{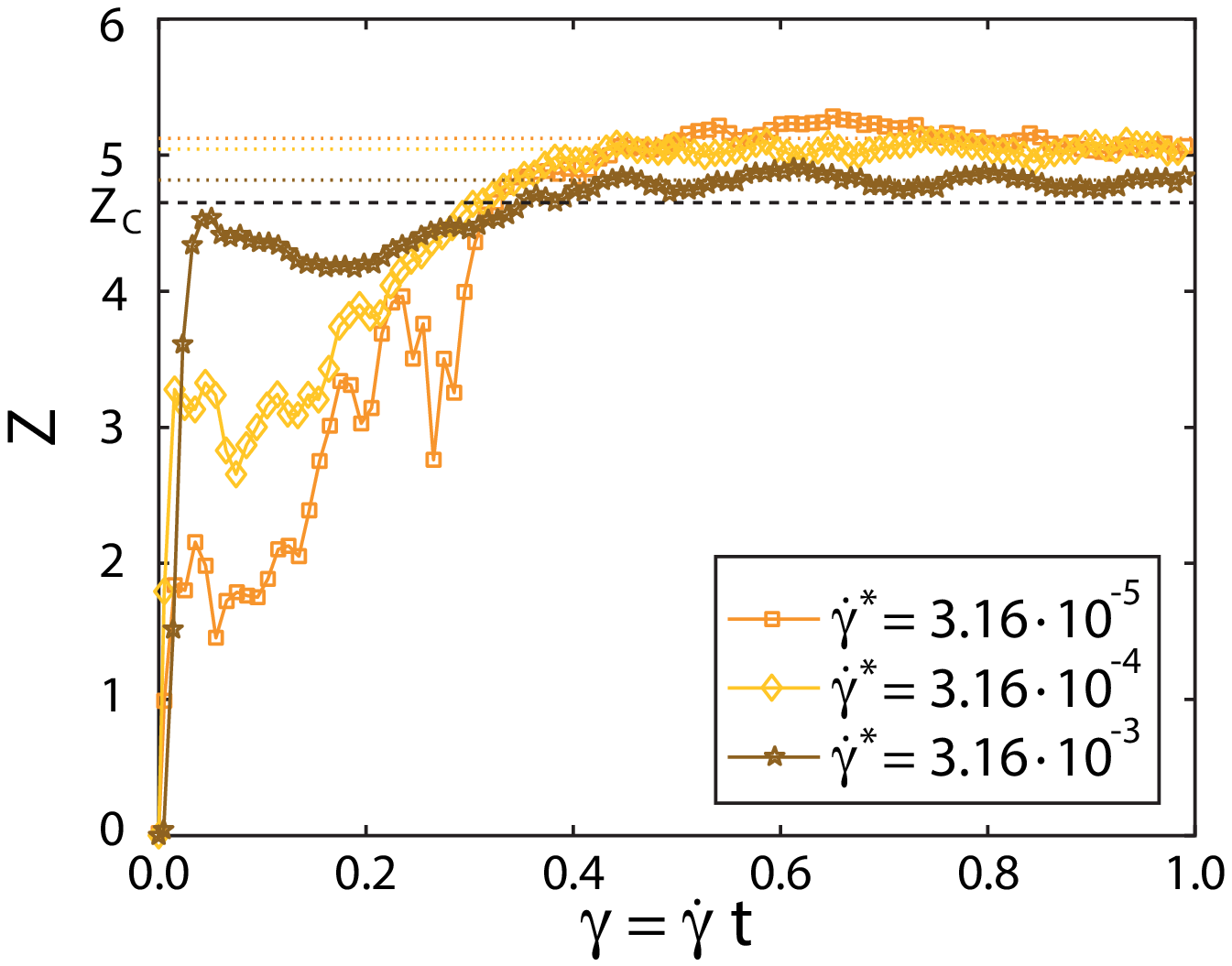}}
\subfigure[\label{nu060_stdC}]%
{\includegraphics[width=0.4\textwidth]{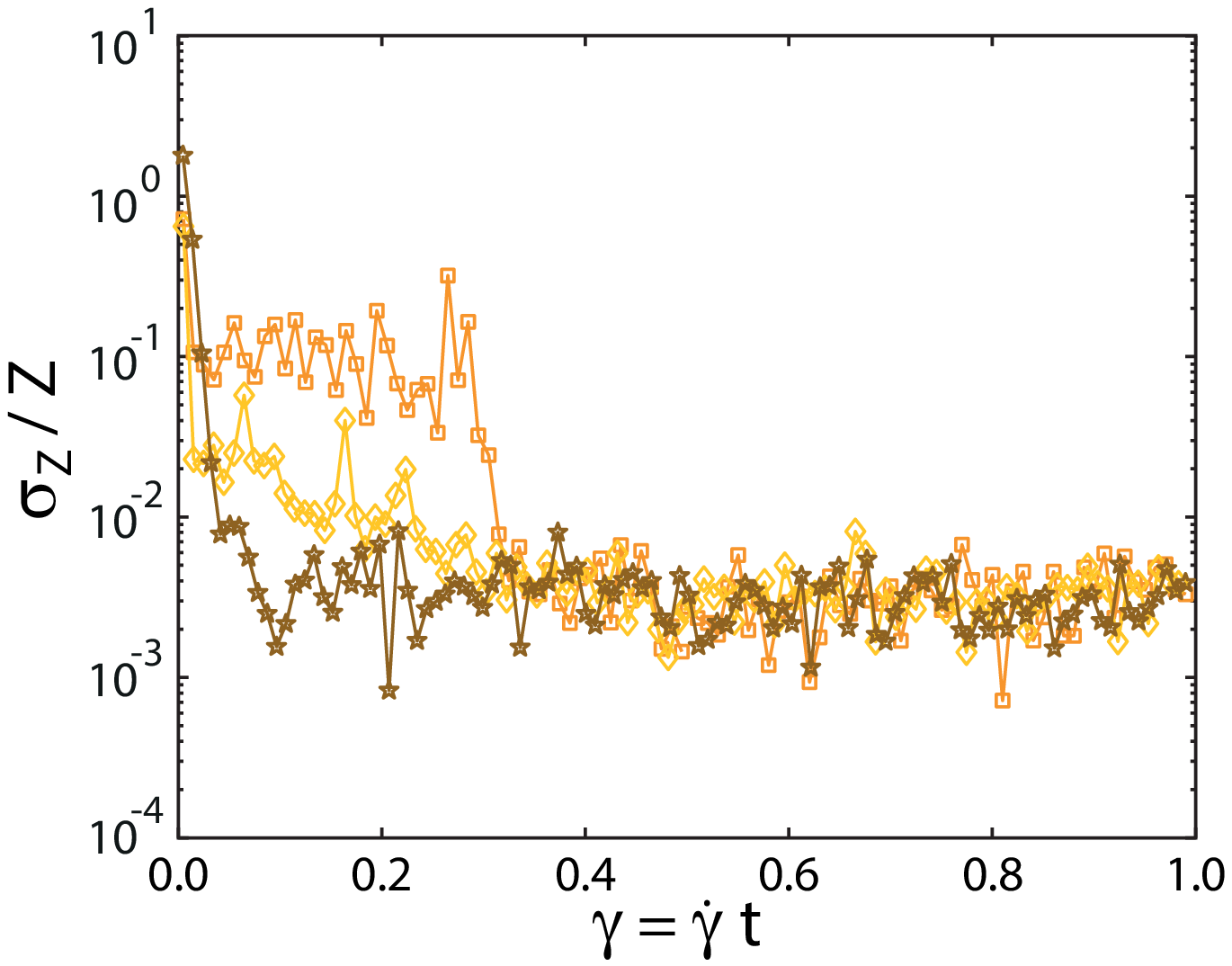}}
\caption{Same as in \Fig\ref{nu059_CstdC}, but when $\nu = 0.60$.}
\label{nu060_CstdC}
\end{figure}
%
%
\begin{figure}[!h]
\centering
\subfigure[\label{nu062_C}]%
{\includegraphics[width=0.4\textwidth]{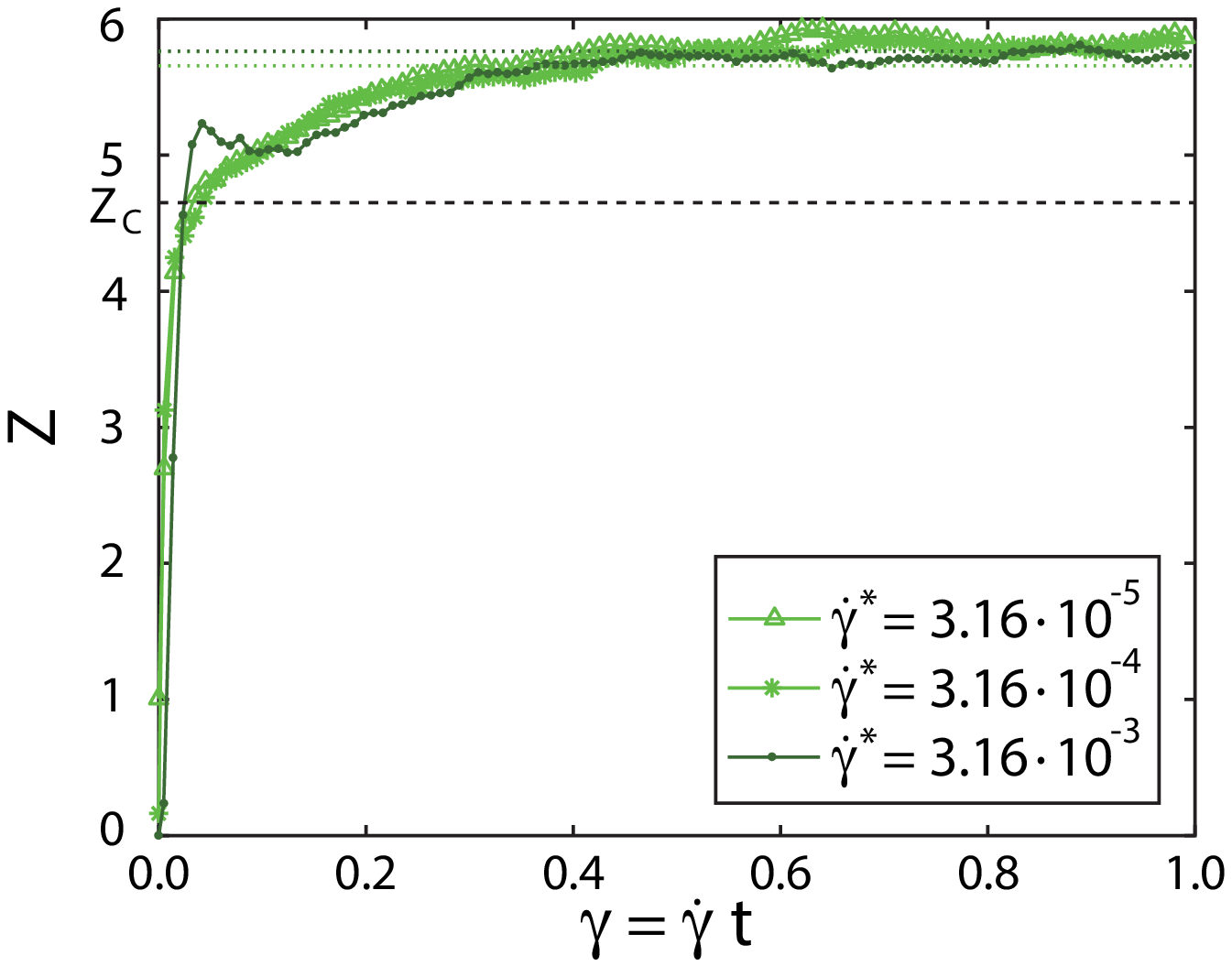}}
\subfigure[\label{nu062_stdC}]%
{\includegraphics[width=0.4\textwidth]{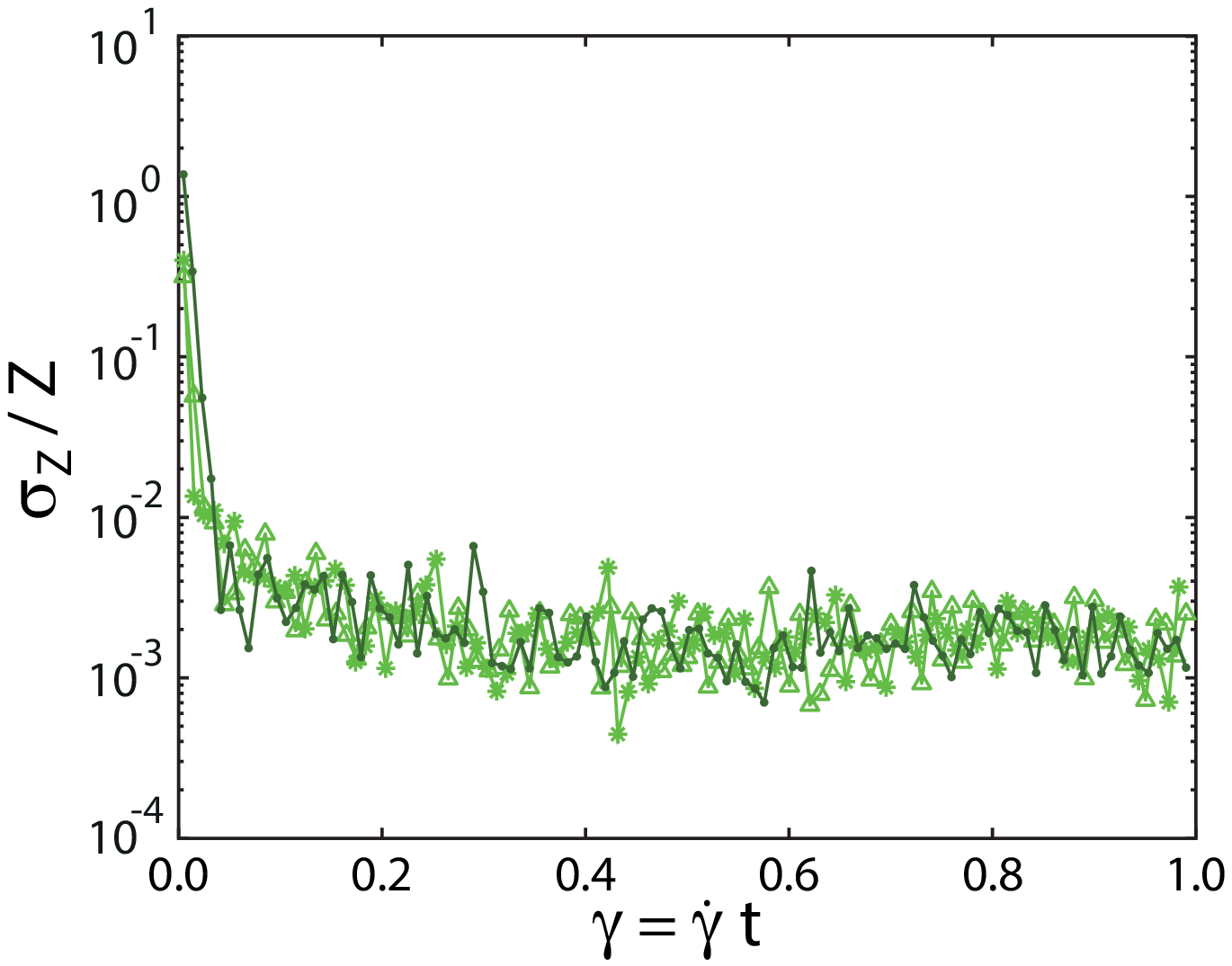}}
\caption{Same as in \Fig\ref{nu059_CstdC}, but when $\nu = 0.62$.}
\label{nu062_CstdC}
\end{figure}
%
%
\begin{figure}[!h]
\centering
\subfigure[\label{nu059_Csol}]%
{\includegraphics[width=0.4\textwidth]{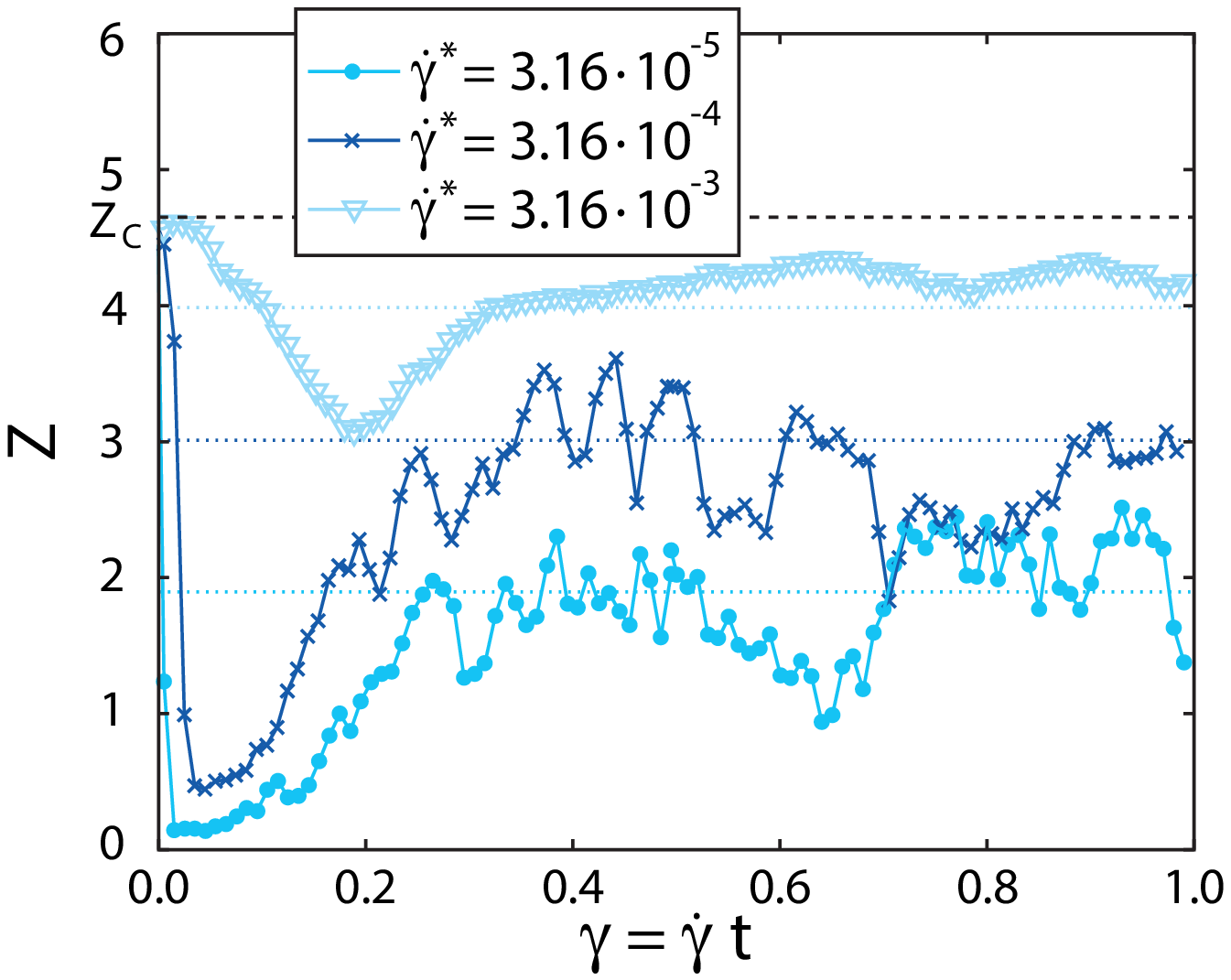}}
\subfigure[\label{nu060_Csol}]%
{\includegraphics[width=0.4\textwidth]{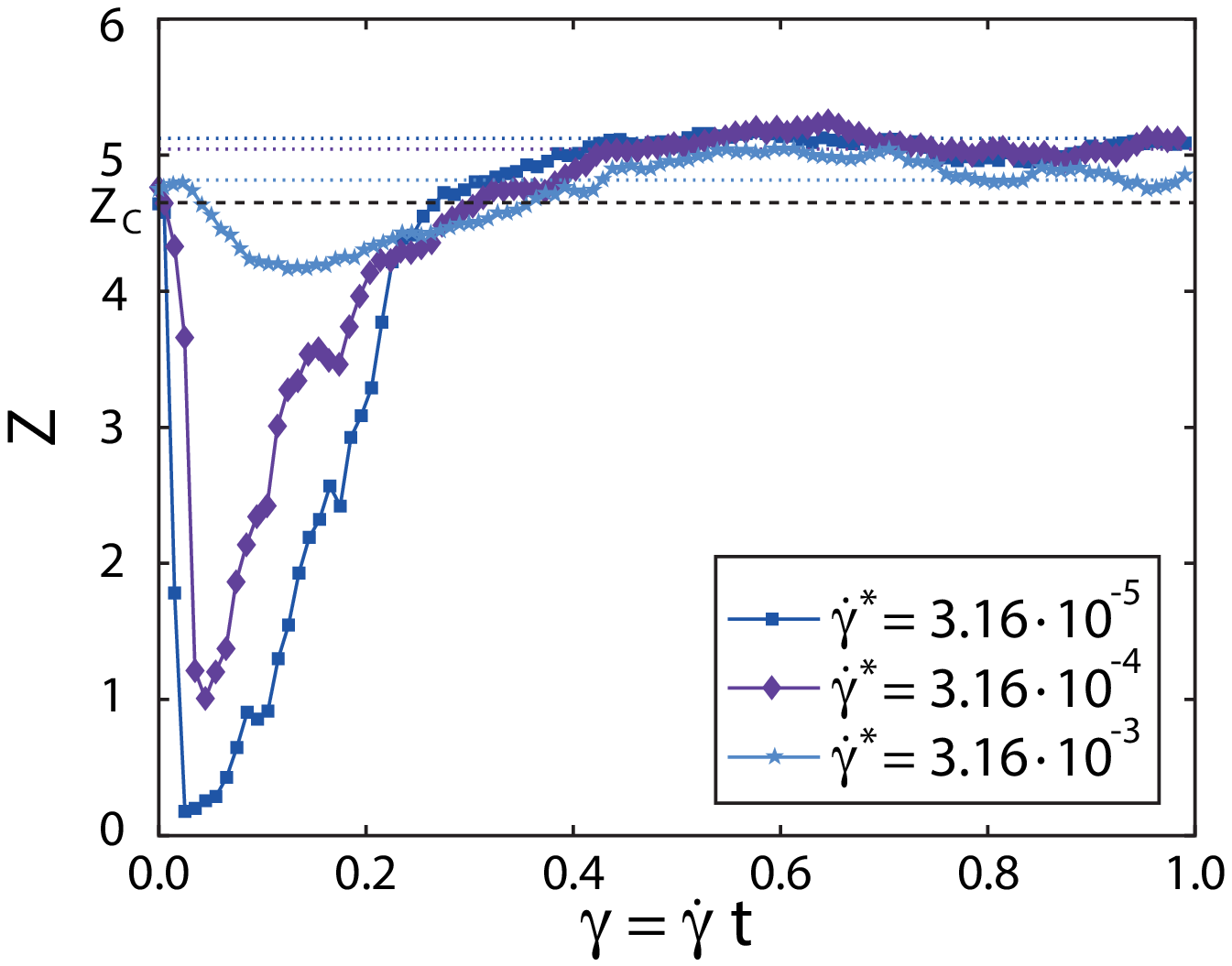}}\\
\subfigure[\label{nu062_Csol}]%
{\includegraphics[width=0.4\textwidth]{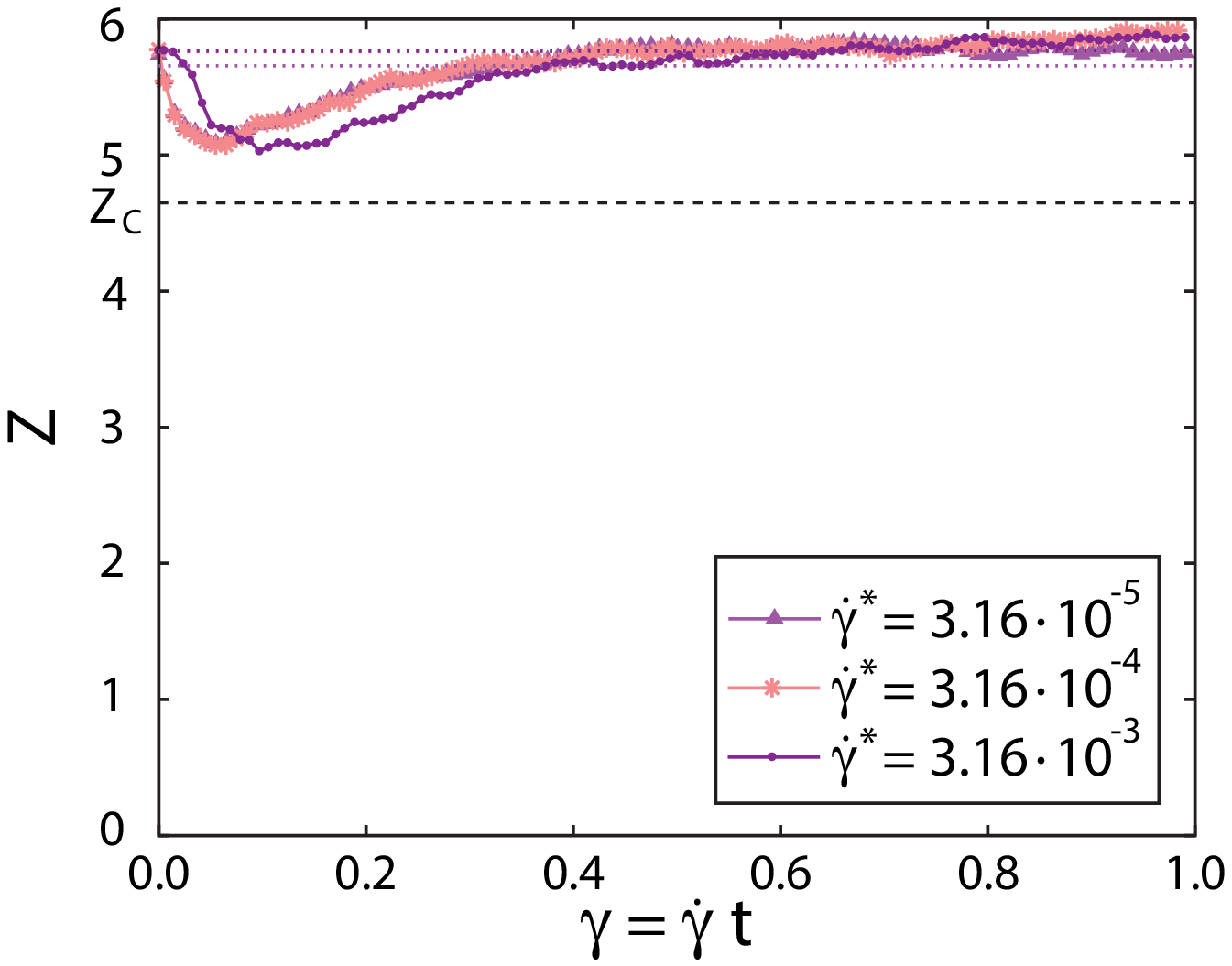}}
\caption{Evolution of the coordination number for different values of the dimensionless shear rate, when (a) $\nu = 0.59$, (b) $\nu = 0.60$,  (c) $\nu = 0.62$, and the specimen is, initially, an isotropic random packing, with $p_0^* \approx 2\cdot 10^{-3}$. Steady state values are denoted with dotted lines; the dashed lines represent the critical coordination number.}
\label{nu_Csol}
\end{figure}
%
The influence of the dimensionless shear rate on the evolution of the coordination number and its fluctuations are illustrated in \Figs\ref{nu059_CstdC}-\ref{nu062_CstdC} for the three volume fractions investigated. These figures refer to the case in which initially isotropic athermal gases are sheared. Three values of the dimensionless shear rate are applied. Large values of $\dot\gamma^*$ are equivalent to either small particle stiffness $k_n$, i.e. soft particles, or rapid shearing. As is shown in \Figs\ref{nu059_CstdC}-\ref{nu062_CstdC}, the fluctuations of the coordination number are strongly affected by the dimensionless shear rate if $Z<Z_C$ and decrease for increasing $\dot\gamma^*$. Conversely, the scaled fluctuations of the coordination number (and also of the pressure, not shown here for brevity) are rate-independent when $Z>Z_C$, reinforcing the idea that the fluid-solid transition takes place at $Z=Z_C$.  
The same effect is shown in \Fig\ref{nu_Csol} where the cases of initially, isotropic random packings, with $p_0^* \approx 2\cdot 10^{-3}$, are considered for the three volume fractions. It is evident that the initial drop in $Z$ is severely affected by ${\dot\gamma}^*$ when $\nu = 0.59$ and $0.60$ (\Fig\ref{nu_Csol}a and b). Conversely, such a dependence seems to disappear when $\nu =0.62$ (\Fig\ref{nu_Csol}c).\\

\begin{figure}[!h]
\centering
\subfigure[\label{pstar_C}]%
{\includegraphics[width=0.4\textwidth]{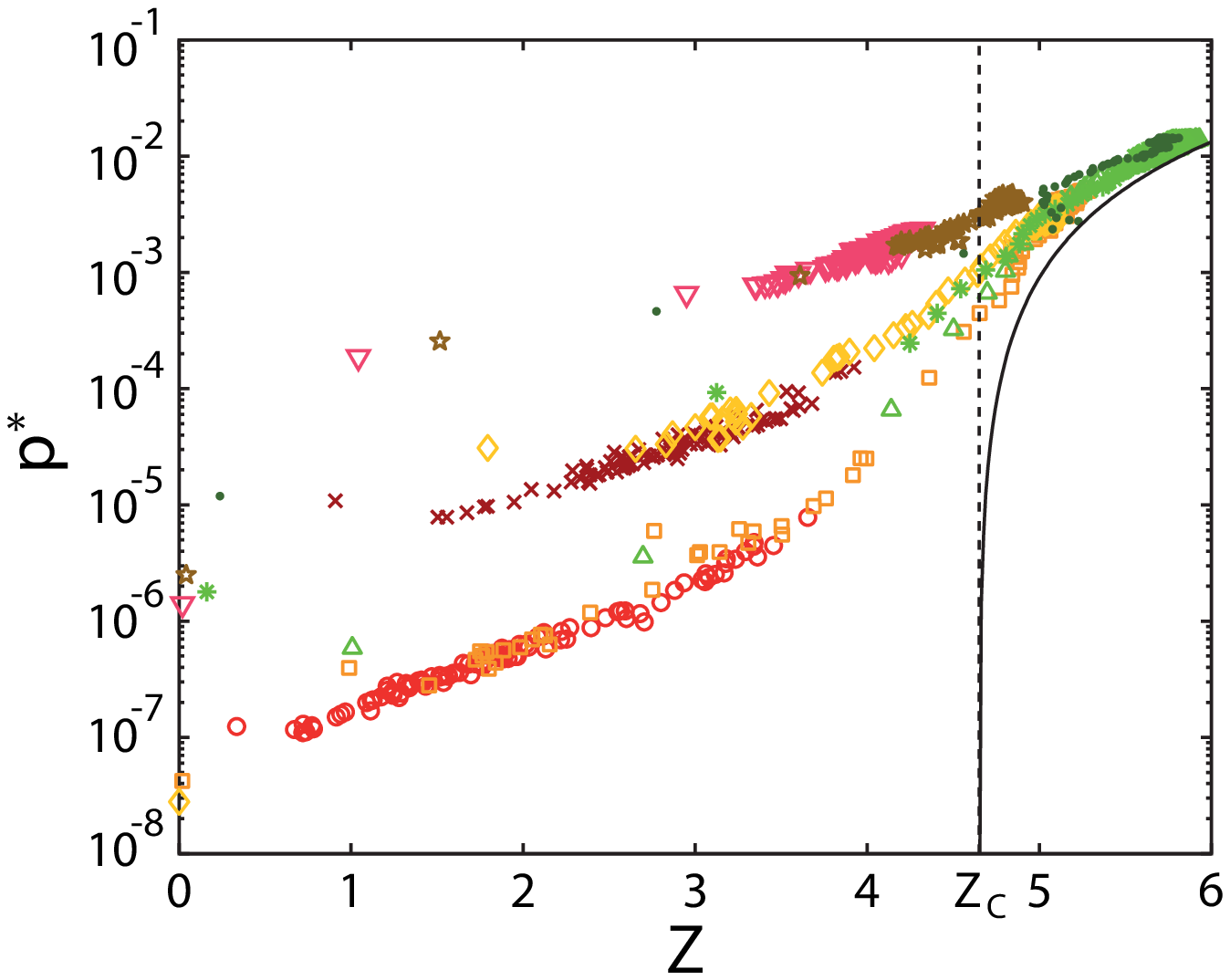}}
\subfigure[\label{Tstar_C}]%
{\includegraphics[width=0.4\textwidth]{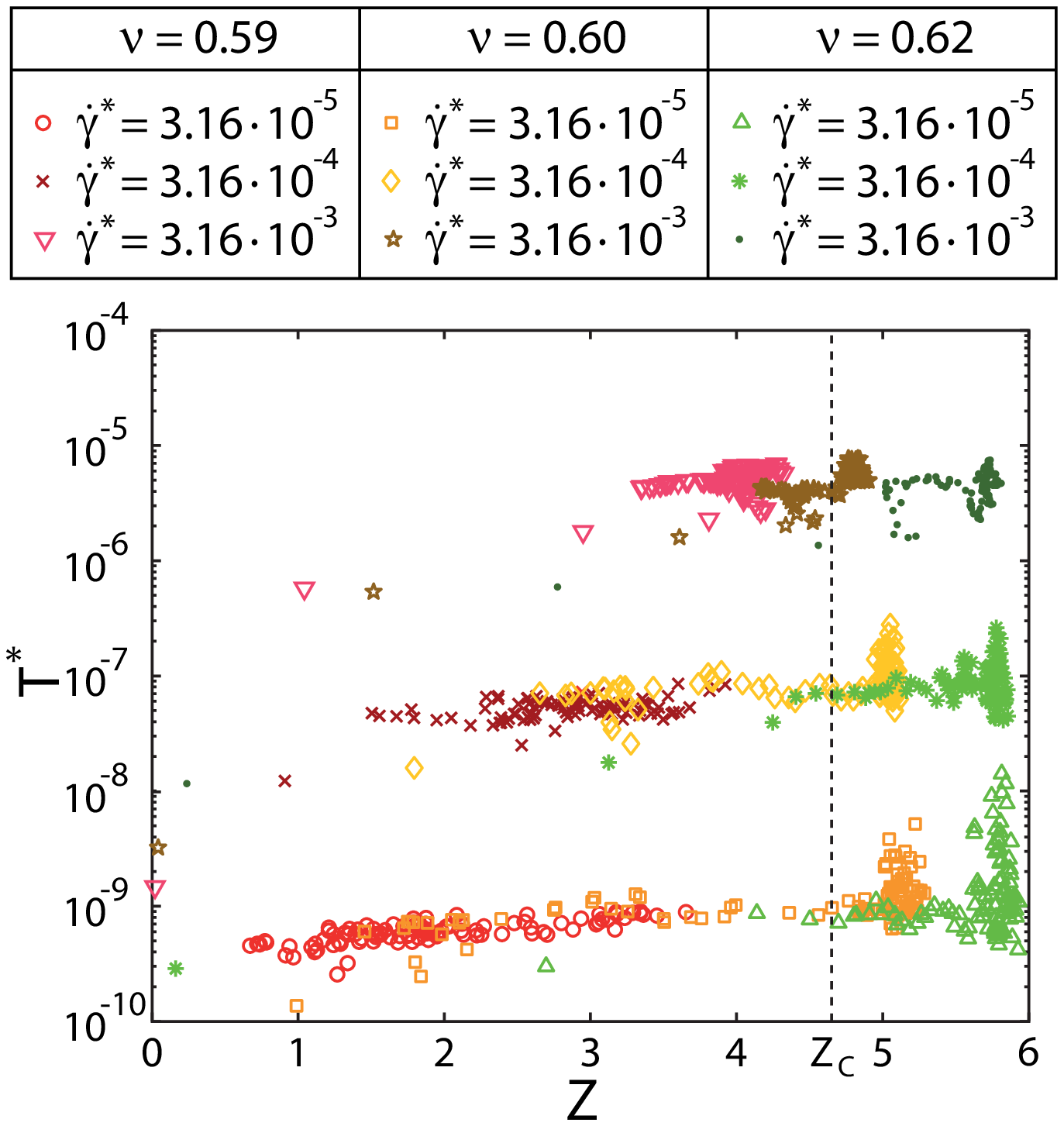}}
\caption{Dimensionless (a) pressure and (b) granular temperature versus the coordination number for different dimensionless shear rates when the specimen is, initially, an isotropic athermal gas. The solid line represents the scaling observed by Sun and Sundaresan \cite{sun2011} in the solid regime.}
\label{pstar}
\end{figure}

Finally, in \Figs\ref{pstar_C} and \ref{Tstar_C}, in case of initially isotropic athermal gas, the dimensionless pressure and granular temperature as functions of the coordination number for different shear rates and volume fractions are shown (all the data plotted in \Figs\ref{nu059_CstdC}-\ref{nu062_CstdC} are here considered). At large values of the coordination number, far from $Z_c$, the measurements of $p^*$ collapse onto the scaling relation observed by Sun and Sundaresan \cite{sun2011} in the solid regime: $p^*\propto (Z-Z_C)^2$ (solid line in \Fig\ref{pstar}a). In contrast, the data do not superimpose when $Z<Z_C$, given the rate-dependency of the pressure in the fluid regime. Near the critical coordination number $Z \approx Z_C$, even in the solid regime, a not negligible rate-dependency is evident. \\
In \Fig\ref{Tstar_C}, the granular temperature is roughly constant at a given dimensionless shear rate, for any value of the coordination number, that is in both fluid and solid conditions. As was already observed in \Fig\ref{fl_gp01_T}, large fluctuations affect $T^*$ when $Z > Z_c$, putting in evidence the collapse of force chains in the solid regime.

%
%

\section{Concluding remarks}

\noindent The fluid-solid transition in granular materials has been investigated by performing discrete numerical simulations of unsteady, homogeneous, shear flows of identical, inelastic, soft spheres at constant volume. A constant shear rate has been instantly applied to either an initially isotropic athermal gas or an initially isotropic random packing. For some combinations of the initial conditions and the constant volume fraction both  fluid-solid and solid-fluid transitions have been observed. 
At slow shearing and/or large stiffness of the particles, the fluid granular assembly exhibits fluctuations in both pressure and coordination number whose amplitude is two orders of magnitude larger with respect to the same material under solid conditions. Analogously, the fluctuations of granular temperature are particularly pronounced in the solid regime, whereas seem to disappear under fluid conditions. This is due to the collapse of force chains and to the transformation of the elastic stored energy into kinetic energy when a network of contact forces develops within the medium which is continuously broken and re-built during the shearing process.
For rapid shearing and/or for small stiffness of the particles, the difference in the fluctuation amplitude is much less evident. However, the fluctuations are rate-independent in the solid state, allowing to clearly identify the phase transition. The fluid-solid transition in unsteady, homogeneous shear flows is characterized by a critical value of the coordination number, independent of the imposed volume fraction and the shear rate. 
Such a critical coordination number coincides with that already defined in the past with reference to steady conditions.
Finally, our results indicate that both the coordination number and the shear rate need to appear in constitutive models for the onset and the arrest of granular flows.

\section*{Acknowledgements}

Dalila Vescovi is supported by a fellowship from Fondazione Fratelli Confalonieri.

\clearpage

\bibliographystyle{plainnat}
\normalem

\end{document}